\documentclass[manuscript,nonacm]{acmart}

\usepackage{longtable}
\usepackage{subcaption}
\usepackage{wrapfig}
\usepackage{listings}

\newif\ifshownotes
\shownotestrue
\relax

\newcounter{challenge}
\renewcommand{\thechallenge}{C\arabic{challenge}}
\newcommand{\challengehead}[2]{%
  \refstepcounter{challenge}%
  \label{ch:#2}%
  \noindent\textbf{\thechallenge. #1}
}

\newcounter{goal}
\renewcommand{\thegoal}{G\arabic{goal}}
\newcommand{\goalhead}[2]{%
  \refstepcounter{goal}%
  \label{goal:#2}%
  \noindent\textbf{\thegoal. #1}
}

\begin{document}

\title{PatchLens: Adapting Mechanistic Interpretability Techniques for Non-Expert Audiences}

\author{Gwenyth Lincroft}
\email{lincroft.g@northeastern.edu}
\orcid{https://orcid.org/0009-0006-8694-3422}
\affiliation{%
  \institution{Northeastern University}
  \city{Boston}
  \state{Massachusetts}
  \country{USA}
}

\author{Yik Su Chan}
\affiliation{%
  \institution{Northeastern University}
  \city{Boston}
  \state{Massachusetts}
  \country{USA}
}

\author{Adam Belfki}
\affiliation{%
  \institution{Northeastern University}
  \city{Boston}
  \state{Massachusetts}
  \country{USA}
}

\author{Emma Bortz}
\affiliation{%
  \institution{Northeastern University}
  \city{Boston}
  \state{Massachusetts}
  \country{USA}
}

\author{Jonathan Bell}
\email{j.bell@northeastern.edu}
\orcid{}
\affiliation{%
  \institution{Northeastern University}
  \city{Boston}
  \state{Massachusetts}
  \country{USA}
}

\begin{abstract}
A chat interface presents a reply as finished text, concealing a model that predicts at every position and reruns.
Mechanistic interpretability can show how a large language model produces text, but its tools are built for researchers.
We present PatchLens, a browser interface that runs logit lens and activation patching on remotely served research-scale models.
Each layer's state appears as the tokens it comes closest to, laid out by layer and input position; an intervention is a drag from one cell onto another.
The tokens a learner reads are the tokens they manipulate.
In a between-subjects study with 61 non-experts, the comparison against Transformer Explainer was inconclusive on post-test knowledge.
PatchLens participants wrote their own prompts (21 of 32 against 1 of 29), reported lower mental demand, and located their difficulty in the interface rather than the material.
We read the null as evidence about what such an instrument measures.
\end{abstract}

\begin{teaserfigure}
  \centering
  \includegraphics[width=\textwidth]{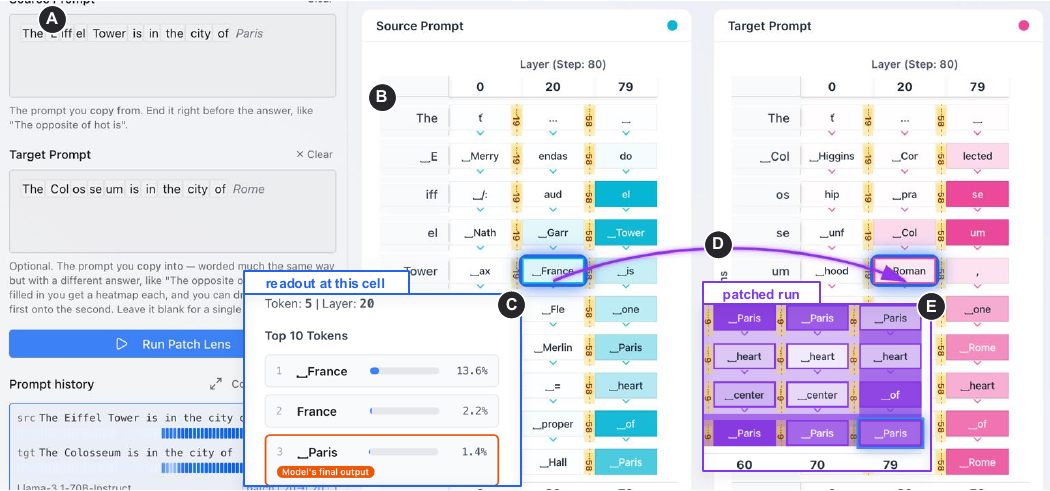}
  \caption{PatchLens brings mechanistic interpretability interventions to
    non-expert users, in the browser, on research-scale open-weight models.
    Here activation patching tests which layers are causally relevant to the
    target prompt's completion, ``The Colosseum is in the city of'' with
    \emph{Rome}.
    (A)~The learner writes a source prompt and a minimally different target
    prompt. (B)~Logit lens renders the residual stream as readable tokens on
    the lens grid: each row an input token position, each column a layer, each
    cell the tokens that layer's intermediate state comes closest to.
    (C)~Clicking a cell opens its ranked readouts; the layer-20 state at
    \emph{Tower} ranks \emph{France} first and \emph{Paris} third.
    (D)~Dragging that cell onto the target at the same layer performs the
    patch. (E)~The target, which predicted \emph{Rome} on its own, now
    predicts \emph{Paris}.}
  \Description{A screenshot of the PatchLens web interface. On the left, a
    prompt editor holds two prompts: "The Eiffel Tower is in the city of"
    and "The Colosseum is in the city of". In the center and right, two
    grids of word predictions, one per prompt, with rows labeled by input
    token and columns labeled by layer. A purple arrow runs from a cell in
    the first grid to the corresponding cell in the second. Two panels
    overlay the grids: one lists the ranked word predictions inside the
    dragged cell, the other shows a fragment of the patched grid in which
    the final prediction reads Paris.}
  \label{fig:teaser}
\end{teaserfigure}

\maketitle

\section{Introduction}
\label{sec:intro}
LLMs produce text that reads as confident and knowledgeable, inviting users to interpret outputs as retrieval, reasoning, or even thinking, when the process is none of these \cite{shanahan_talking_2024}.
Users consequently build mental models from the system's surface behavior and struggle to predict or diagnose its failures \cite{zamfirescu-pereira_why_2023}.

\citet{cho_transformer_2026} found that Transformer Explainer improved non-expert understanding of model components.
Interactive explanations and video walkthroughs of this kind stop at architecture: they show what a transformer is built from, not how a prediction was formed from a given input.
A non-expert who watches a video walkthrough \cite{sanderson_transformers_2024} and then wants to trace a specific prompt through a model, layer by layer, has nowhere to go.
Knowing that a transformer has attention heads and multilayer perceptrons does not, on its own, give the user intuition into \textit{what the model is doing right now with their input}.

The perceivable boundaries between a system's parts are its \emph{seams} \cite{weiser_creating_1994}, which the seamful design tradition reveals selectively rather than removes \cite{chalmers_seamful_2003, chalmers_seamful_2004}.
Later work carries the stance into explainable AI \cite{ehsan_seamful_2024}.
A chat interface conceals a seam of this kind. It presents a reply as one block of finished text, while the process behind it computes a prediction at every input position and runs the model again for every token.

Mechanistic interpretability describes how model output is generated by reverse engineering a model from its fundamental components \cite{elhage_mathematical_2021}.
Two technique families show how the output token is produced. Logit lens \cite{nostalgebraist_interpreting_2020, belrose_eliciting_2025} projects a model's intermediate states back into its own vocabulary, so each layer's state can be read as tokens. Activation patching \cite{meng_locating_2023, vig_investigating_2020} moves a state from one run into another and reports the effect on the output.
The tools that support this work, though, are built for machine learning (ML) researchers: none let non-experts directly manipulate interpretability interventions on research-scale open-weight LLMs.
Educational tools do not expose interventions---research tools assume a background non-experts lack.

We present PatchLens, a browser-based interface that brings logit lens and activation patching to people without an ML research background, on the open-weight models interpretability researchers study.
Models are served remotely on the National Deep Inference Fabric (NDIF), which opens research-scale inference to projects such as ours.
We evaluated PatchLens against Transformer Explainer \cite{cho_transformer_2026}, the most widely used interactive resource for learning transformer architecture, in a between-subjects study with 61 non-experts.

We contribute:
\begin{enumerate}
    \item \textbf{PatchLens}, the first browser-based tool that exposes two mechanistic interpretability interventions, logit lens and activation patching, on research-scale LLMs to non-expert users.
    Together they let a learner ask both what a model represents and what it actually uses, on prompts they write themselves rather than curated examples.

    \item \textbf{Interaction techniques for non-expert exploration of LLM internals.}
    Logit lens is the primary interaction surface: because every layer's hidden state is projected back into the vocabulary, an intervention manipulates readable tokens rather than edits a hidden representation, and its result is reported in the same frame.
    We identify five challenges in adapting mechanistic interpretability research tools for non-expert use, the five design goals that follow, and the interventions we built in response.

    \item \textbf{A controlled comparison against the most widely used tool of its kind} with 61 non-experts, reporting what participants learned, what they chose to run when the prompt was theirs, and how the two conditions were experienced.
    The study could not detect a post-test knowledge difference of the size it was powered to find, so we report neither an advantage nor an equivalence. Participants wrote and ran their own prompts, reported lower mental demand than Transformer Explainer participants, and located their difficulty in the interface rather than the material.
\end{enumerate}

\section{Background}
\label{sec:background}
PatchLens's two techniques operate on the internal state a transformer produces as it processes input.
Mechanistic interpretability describes that computation: which internal components represent what, and which causally drive specific behaviors \cite{saphra_mechanistic_2024, olah_zoom_2020}.
Its techniques divide into two groups: \textit{observational} techniques read internal state without altering it, showing \textit{what the model represents}, and \textit{interventional} techniques modify an activation and measure how the output changes, showing \textit{what the model uses}.
This section introduces that state, then logit lens (observational) and activation patching (interventional).

\subsection{Transformer Text Generation}
\label{sec:transformers}

\textit{How does a language model produce text?}

A transformer-based LLM first splits input text into \emph{tokens}, the units of text the model reads.
Llama-3.1-70B-Instruct, for instance, requires four tokens for ``Colosseum,'' and each model tokenizes differently.
Each token is mapped to a high-dimensional vector, and that vector is refined by a stack of \emph{transformer blocks} (Fig.~\ref{fig:logitlens}).
The vector held at a given layer and token position is an \emph{activation}.\footnote{We use the term throughout for the state at a layer and position: what logit lens reads and what PatchLens patches.}
The activations at one token position, one per layer, form the \emph{residual stream} at that position: the model's running state for that token.

Attention, one of two components of a transformer block (the other is a multilayer perceptron), lets each position read from the residual streams of other positions.
That access is \emph{causally masked}: a position may read from itself and from \emph{earlier} positions, but never from later ones.
Which internal states were available to produce a given one is fixed by the architecture rather than the input.

A final \emph{unembedding} step projects the final layer's state into a score over the model's vocabulary, normalized into a probability distribution from which the output token is taken.
This step applies at every position, so a prediction is computed at each input position rather than only at the last.
Producing more than one token requires appending that token to the input and running the model again.
Each run produces one token, and has available only the model's pre-trained weights and the text in the prompt.

\subsection{Logit Lens}
\label{sec:LogitLens}

\textit{If the model had to commit to a prediction right now---at this layer, before processing is complete---what would it choose?}

\begin{figure}
    \centering
    \includegraphics[width=0.8\linewidth]{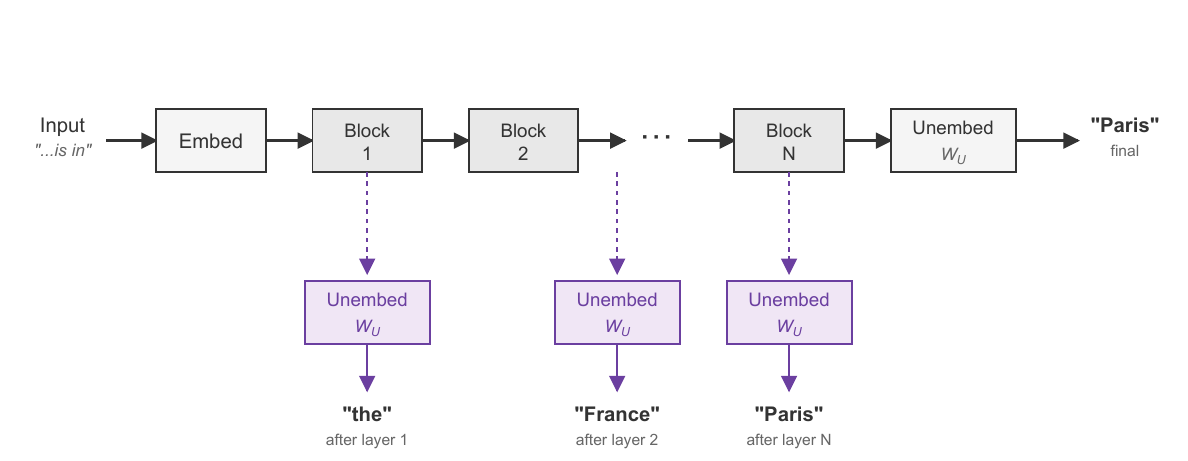}
    \caption{Logit lens applied to intermediate activations of an LLM, in purple, using the same unembedding matrix $W_U$ that produces the model's output. The diagram reads the final input position, from which the next token is predicted; in practice any position can be read. The token predictions shown are illustrative.}
    \label{fig:logitlens}
    \Description{A horizontal flow diagram of a transformer language model. From left to right: an input prompt feeds into an embedding block, which connects to a sequence of transformer blocks labeled Block 1, Block 2, ellipsis, Block N. After Block N, a final unembedding block labeled W subscript U produces the output token "Paris". Below the main flow, three purple branches descend from Block 1, Block 2, and Block N. Each branch passes through its own unembedding box labeled W subscript U and produces a token prediction: "the" after layer 1, "France" after layer 2, and "Paris" after layer N. The three intermediate unembedding operations and the final one all use the same W subscript U.}
\end{figure}

A transformer builds its prediction gradually, and the unembedding step normally runs only at the final layer.
Logit lens applies it at each intermediate layer as well, producing a vocabulary-space estimate per layer (Fig.~\ref{fig:logitlens}).
Early layers often produce surface-level or apparently unrelated continuations, and later layers converge on the final answer \cite{nostalgebraist_interpreting_2020}.
Results are conventionally presented as a heatmap over layers and token positions, encoding the most probable token and one scalar (probability, rank, or entropy) per cell.

Logit lens approximates: it shows what information is present in an activation at a given layer, not what the model will say or what it uses internally.
A token that ranks highest at layer 10 may be overwritten by later layers, and information the model relies on downstream may not appear as a top-ranked token at all.

\subsection{Activation Patching}
\label{sec:ActPatch}

\textit{Which internal components, if changed, would change the model's output?}

\begin{figure}
    \centering
    \includegraphics[width=0.8\linewidth]{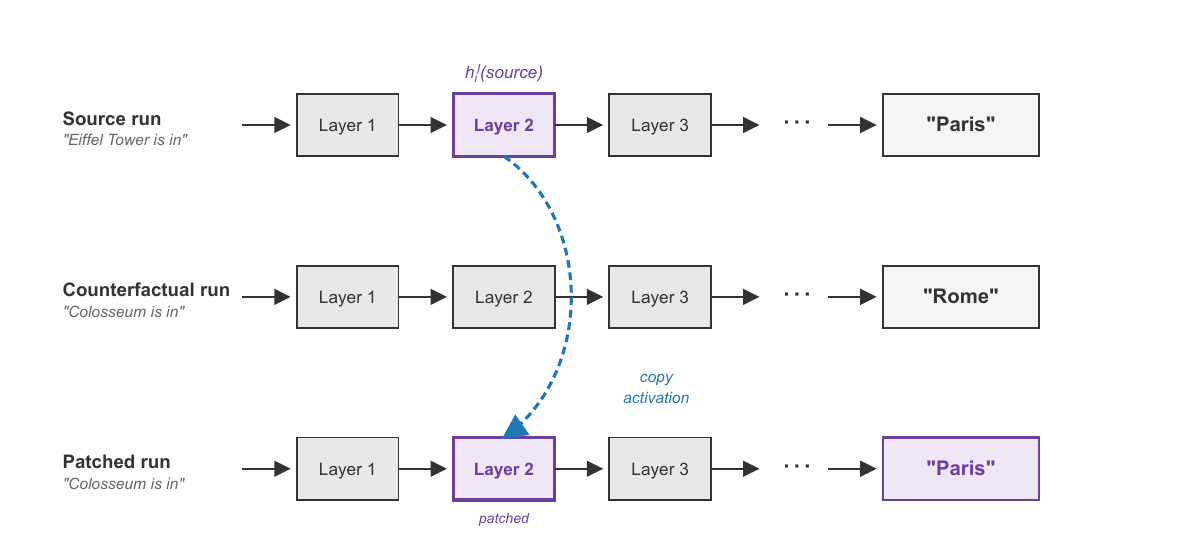}
    \caption{Activation patching as a causal intervention, with the substituted activation and the runs it affects in purple. Replacing $h_i^l$, the activation at layer $l$ and position $i$, shifts the output from ``Rome'' to ``Paris'', so that activation was causally sufficient to produce the source behavior in the target context.}
    \label{fig:activation_patching}
    \Description{A diagram showing three parallel forward passes of a transformer language model, arranged in three horizontal rows. The top row, labeled "Source run" with prompt "Eiffel Tower is in", passes through layers labeled Layer 1, Layer 2, Layer 3, ellipsis, and produces the output "Paris". The Layer 2 box in this row is highlighted in purple and labeled h subscript i superscript l (source) to indicate it is the activation being extracted. The middle row, labeled "Target run" with prompt "Colosseum is in", has the same layer structure and produces the output "Rome"; no layer is highlighted. The bottom row, labeled "Patched run", uses the same target prompt "Colosseum is in" but its Layer 2 box is highlighted in purple and marked "patched". A dashed blue curve labeled "copy activation" connects the highlighted Layer 2 box of the source run to the highlighted Layer 2 box of the patched run, indicating the activation is being transferred from one run to the other. The patched run's output is "Paris", also rendered in purple to signal that it has shifted from the target run's behavior ("Rome") to match the source behavior ("Paris") as a result of the substitution.}
\end{figure}

Where logit lens reads internal state, activation patching (Fig.~\ref{fig:activation_patching}) intervenes on it.
Activation patching applies causal mediation analysis \cite{pearl_direct_2001}: setting an intermediate variable, here an activation, to the value it would have taken under a different input, then observing what changes in the output.

The experiment uses two prompts and three forward passes.
A \emph{source} prompt produces the behavior of interest, here ``The Eiffel Tower is in'' $\rightarrow$ ``Paris''.
A minimally different \emph{target} prompt produces a different one: ``The Colosseum is in'' $\rightarrow$ ``Rome''.
To test whether an activation carries the information distinguishing the two behaviors, a third pass runs the target prompt with the activation at one layer and position replaced by the source run's activation there.

If the patched output shifts toward the source behavior (``Paris''), the activation was causally sufficient to produce that behavior in the target context; if it is unchanged, the relevant information was encoded elsewhere.
Results are conventionally reported as line plots of how the patched run's output varies with patch position in layers.
In research practice a patching experiment sweeps across layers, positions, and prompt pairs rather than reporting a single patch \cite{meng_locating_2023, wang_interpretability_2023}.

\section{Related Work}
\label{sec:relwork}
Three questions organize the prior work: what a revealed mechanism must look like (\S\ref{subsec:seamful}), what non-experts have been taught about LLMs and on what models (\S\ref{subsec:tools}), and who can run an intervention (\S\ref{subsec:mitools}).

\subsection{Seamful Design and What a System Reveals}
\label{subsec:seamful}

\citet{weiser_creating_1994} advocated systems with ``beautiful seams.''
\citet{chalmers_seamful_2003} developed that into a design stance that questions whether seamless integration is necessarily a requirement and treats revealing and using seams as deliberate policy.
\citet{inman_beautiful_2019} reject the reading that casts seamful and seamless design as rivals, on the observation that no system reveals everything.
The design question is therefore not whether to reveal a mechanism but which of its boundaries to open and which to leave closed.
Concealment is as much a design act as revelation.

What constrains that choice is what the user is meant to do once the mechanism is open.
\citet{chalmers_seamful_2004} supply the prescription this paper designs against: a user can reveal a seam when the task is to understand or change the mechanism beneath it, through a representation that is robust, simple, and flexibly manipulable.
The three criteria are demanding, and may not be separable: a seam revealed so that someone can change what lies behind it has to be exposed through a surface they can act on directly, in terms they can already read.

Applied to AI, the stance has so far been taken up at the scale of deployment.
\citet{ehsan_seamful_2024} treat a seam as a mismatch between an AI system's design assumptions and the reality of its deployment context, located across the system's lifecycle rather than inside the model, and ask which of those mismatches its stakeholders should be shown.
The mismatch this paper addresses is the same kind in a different place: the assumption is that a single appended reply adequately accounts for what produced it, and it holds between an interface and the mechanism beneath it rather than between a model and its context of use.

\subsection{Interactive Tools for Learning About LLMs}
\label{subsec:tools}

Tools that teach non-experts about LLMs differ in where the model runs, and that choice determines both the scale of the model a learner meets and how much of it they control.

\subsubsection{Interactive tools trade model scale against learner control.}
\label{subsubsec:scaletrade}
Video and static walkthroughs of LLMs have reached millions of non-expert learners (e.g., 3Blue1Brown \cite{sanderson_transformers_2024}), but a linear format cannot let a learner act on the process it describes.
Running the model in the browser buys live inference on learner-chosen input at the cost of scale, and the three tools that make this trade differ in what they do with it.
Transformer Explainer \cite{cho_transformer_2024, cho_transformer_2026} runs a live GPT-2 whose input text and sampling hyperparameters non-experts manipulate against token-centric flow visualizations of the architecture, attention heads, and output distribution.
Its users answered conceptual questions more accurately and reported higher understanding than users of static blog posts or videos.
That evaluation identified custom input as what learners valued.
LLM Visualization \cite{bycroft_llm_nodate} instead renders every matrix multiplication of a working nano-GPT in a 3D walkthrough a learner can zoom into, and fixes the prompt to afford that depth.
\citet{hernandez_vit-explainer_2026} pair patch-level attention overlays with a vision-adapted logit lens that traces class predictions across a Vision Transformer's layers, the first interactive learning tool to show internal representations alongside architecture explanation, though on a pedagogically reduced model.

Abandoning live inference relaxes the scale constraint and removes learner control altogether.
AnimatedLLM \cite{kasner_animatedllm_2026} animates tokenization, attention, and autoregressive generation from traces pre-computed over several open models, so a fully client-side application covers frontier-scale models across multiple languages and adds a view of the pre-training objective.
Nothing in it is run by the learner.
TransforLearn \cite{gao_transforlearn_2024} shows that the tradeoffs are not only about scale: its clickable component-level views expose the mathematical operations inside each layer of an encoder-decoder architecture without a continuous data-flow representation.
Its participants still answered most conceptual questions more accurately than a blog tutorial's readers, several of them significantly faster: early evidence that component-level drill-down supports learning.

No tool we know of holds live inference, learner-chosen input, and research-scale models together, because the browser forces the trade.
PatchLens keeps all three by serving the model remotely, at the cost of a dependency not every project can assume.
These tools also favor the architecture over the prediction: how a particular prediction was formed, what the model weighed and discarded, and what follows from altering an internal state all stay on the far side of the interface.

\subsubsection{Surface-level interactivity may be insufficient for teaching about LLMs.}
\label{subsubsec:surfacelevel}
\citet{zhao_thinking_2025} compared four interactive explorables against static PDF versions of the same content and a no-tutorial control, assigning each participant to one of four topics: large language models, fairness, diversity, and data representation.
The LLM condition used ``Fill in the Blank'' \cite{pearce_what_2021}, which lets users mask a word in a sentence, inspect a masked language model's candidates for that position, and watch them shift as they alter the surrounding words.
That surface invites learners to form and test hypotheses through controlled changes to the input.
Interactive exploration improved understanding of the three non-LLM topics, but performance on the LLM topic declined in every condition, including the no-tutorial control.
The authors suggest that language model behavior is harder to learn than topics grounded in concrete frameworks, and that generative AI may require scaffolding such as guided examples or case-based reasoning.

\subsection{Mechanistic Interpretability \& User Understanding}
\label{subsec:mitools}

Mechanistic interpretability has produced both the techniques PatchLens exposes and the tools that run them, yet almost all evidence about how users read interpretability output comes from elsewhere.

\subsubsection{Mechanistic interpretability tools assume technical expertise.}
Tools for making a model's internal computation inspectable include visual analytics systems for attention \cite{yeh_attentionviz_2024}, feature dictionaries \cite{lin_neuronpedia_2023}, circuit-level tracing \cite{ameisen_circuit_2025}, and observability platforms \cite{meng_monitor_2024}.
All have enabled substantial progress on what large language models represent and how they compute, and all are designed for researchers.
Their interfaces assume familiarity with the computational machinery, and their outputs are not directly human-readable.
The supporting libraries are more demanding still: TransformerLens \cite{nanda_transformerlens_2022} and NNsight \cite{fiotto-kaufman_nnsight_2024} require familiarity with how LLMs are implemented in PyTorch.
NDIF \cite{fiotto-kaufman_nnsight_2024} reduces the engineering burden of running research-scale models by separating experiment specification from model deployment, so interventions execute as deferred computation graphs on shared remote infrastructure.
It still stops at the boundary of programming expertise, which PatchLens removes rather than lowers by replacing code with direct manipulation.

\subsubsection{Non-expert engagement with mechanistic interpretability has been limited to curated exploration.}
The closest non-expert-facing engagement with mechanistic interpretability is the Patchscopes Explorable \cite{hussein_can_2024}.
Patchscopes generalizes activation patching, decoding information from a patched representation using an inspection prompt.
The explorable walks readers through three case studies in entity resolution, attribute extraction, and multi-hop reasoning correction, with interactive visualizations of predefined source-target patching configurations.
This establishes that mechanistic interpretability can be communicated to non-experts, but it does so as a curated explorable on fixed examples rather than as a tool readers can run on their own prompts.

\subsubsection{Evidence on user engagement with interpretability comes from non-mechanistic methods.}
Research on user-facing interpretability tools is dominated by techniques that explain individual predictions by quantifying input importance, and that evidence is substantial but mixed, drawn mostly from applications outside LLMs.
At the technique level, saliency maps fail standard sanity checks, often looking similar regardless of whether the model has learned anything \cite{adebayo_sanity_2018}.
At the user level, end-users misinterpret saliency-based explanations systematically, attributing causal meaning to highlighted regions the model does not rely on \cite{schuff_human_2022}.
ML practitioners using these tools in their own workflows over-trust the outputs and apply them beyond what the methods support \cite{kaur_interpreting_2020}.
Effects on trust in LLMs are heterogeneous, with explanations calibrating trust appropriately in some studies \cite{bauer_explained_2023} and inflating it in others \cite{ahn_impact_2024, duro_measuring_2025}.
This literature has not addressed whether non-experts can engage productively with mechanistic explanations.

\section{Formative Study: Novice Interaction with Logit Lens}
\label{sec:formative}
Before designing PatchLens, we ran a formative workshop with thirteen philosophy graduate students, none with a background in ML research. Most had encountered LLMs only as users and met transformers at the summer school.
They worked through a notebook tutorial on logit lens, running Llama-3.1-8B on NDIF~\cite{fiotto-kaufman_nnsight_2024}, then explored a question of their own.
We collected surveys and field notes.
We asked whether and how non-experts can engage productively with a mechanistic interpretability technique for their own research.

\textbf{Non-experts produced mechanistically interesting findings but could not generalize them.}
Nine of the ten who completed it found indications of potentially interesting behavior on their own prompts.
F8, exploring whether gendered stereotypes appear in intermediate layers, wrote the prompt \lstinline|man: woman, king: queen, actor: actress, pilot: |, and found that several ranked \verb|steward| highly before the final layer settled on \verb|co|, presumably for \emph{copilot}.
F5 tested a \lstinline|roses are red| template for rhyme, and observed near-rhyming non-English tokens (\verb|uu|, \verb|aroo|) in intermediate layers before the final rhyming completion.
F8 nonetheless remained ``not sure how I could use this to draw firm conclusions about general trends in model behavior.''

\textbf{Conventional logit lens visualizations did not connect to the model.}
Ten of the 13 rated themselves ``not at all'' or ``slightly'' familiar with heatmaps, and several asked for a more direct, token-by-token rendering of the layer-by-layer trace.
Participants also struggled to map the heatmap onto the residual stream.

\textbf{The barrier was experiment design, not the technique itself.}
The notebook workflow forced participants to read function signatures and chain Python calls even when their conceptual question was simple.
Beyond programming, logit lens works best with a prompt whose continuation is a single predictable token: F5's first rhyme prompt drew a more frequent non-rhyming token, and only a revision produced the intended contrast.
Participants posed questions and wrote single prompts without much trouble, but struggled to design and carry out generalized experiments.
F11 generated more prompt variants than they could read (``I wasn't able to look at all the graphs because there were too many of them'') and asked to separate running the technique from managing its results.

\textbf{Participants wanted to intervene, not just observe.}
F11 described logit lens as ``only a partial view''; F13 noted that the technique ``merely point[s] to interesting behavior.''
Asked what they would want in a tool, several spontaneously described a causal intervention: a way to test whether a representation logit lens reveals is one the model actually uses, which is why PatchLens exposes both techniques.

\textbf{The model under inspection did not match the model participants had in mind.}
Participants were confused that the model predicts a next token at every input position rather than only at the end.
The unplanned discussion took up much of the session.
F3 called the digression ``really, really helpful'' and asked for more on architecture.
Several doubted the technique applied to their own questions, because the workshop ran a base model rather than an assistant that could ``output bodies of text'' (F10).

\textbf{Mechanistic results were difficult for participants to evaluate.}
Asked for their view of interpretability techniques, several declined to hold one or attributed their trust in logit lens to their interaction with the facilitators rather than to anything in its output.
Most reported that their understanding of LLM text generation had expanded, but confidence in explaining that process rose for some and fell for others.
Those whose confidence fell had been the most confident beforehand.
F13, who entered with a CS background, reported ``I think I understand a lot less about how LLMs work based on the logit lens outputs,'' and F9 that ``I am now more aware of how confused I am as to how transformer models work.''
We read this as recalibration rather than loss, because the same participants were producing competent observations.

These observations locate the design space along three axes: how internal state is rendered, what surrounds an intervention before and after it is run, and what a learner can conclude from a mechanistic result.
The exposure acted on participants' model of the system rather than only on their research question, which is what redirected PatchLens toward education.
We develop these axes into five design challenges (\S\ref{sec:challenges}) and five design goals (\S\ref{sec:designgoals}).
The study was small ($N = 13$), exploratory, drawn from a single discipline, and run on a notebook rather than a graphical interface, so we treat it as formative evidence about where the design space lies rather than as an evaluation of a specific tool.

\section{Design Challenges}
\label{sec:challenges}
We identify five design challenges (\ref{ch:unreadable}--\ref{ch:modelgap}) in adapting mechanistic interpretability interventions to LLM education. They arise from the techniques themselves, from prior work on interactive learning tools and non-expert engagement with interpretability, and from direct observation (\S\ref{sec:formative}).
 
\challengehead{Model internals have no human-readable form; default visual idioms presume fluency non-experts lack.}{unreadable}
Logit lens maps activations into the model's vocabulary, but its conventional heatmap presentation (\S\ref{sec:LogitLens}) puts layer depth and token position on axes that formative participants could not connect to the residual stream (\S\ref{sec:formative}).
Diagrammatic explanations of activation patching place the intervention in the residual stream (Fig.~\ref{fig:activation_patching}). The plots that report its results do not, and presume a researcher can reconstruct that context unaided.
Both fail the first two criteria \citet{chalmers_seamful_2004} set for a revealed mechanism (\S\ref{subsec:seamful}): the representation is not \emph{robust}, because the internal state has no legible form of its own, and not \emph{simple}, because the default idioms presume a fluency non-experts lack.
 
\challengehead{An intervention is only one part of an experiment.}{workflow}
Presented as single interventions (\S\ref{sec:LogitLens}, \S\ref{sec:ActPatch}), neither technique carries the experimental practice that surrounds it in research use.
Logit lens needs a prompt whose continuation is a single predictable token, which rules out most naturally written text and makes prompt design iterative. A prompt ending at a sentence boundary yields punctuation, an answer spanning several tokens yields a word fragment, and a trailing space changes which token is produced.
A readable output is not the same as an informative one, which makes logit lens the harder case.
Activation patching compounds these constraints: a researcher must first convert a question into a source and target that differ minimally, produce contrastive behavior, and yield predictable single-token outputs.
After either intervention, the user must make sense of the output: how the result relates to their motivating inquiry, and what to run next.
\citet{de_jong_scientific_1998} identify hypothesis generation, data interpretation, and the design and tracking of experiments as the difficulties learners meet in these environments, where our formative participants struggled as well (\S\ref{sec:formative}).
The measures they propose include domain information on demand and assignments with concrete goals.
 
\challengehead{A result informative to a researcher can be indistinguishable from a malfunction to a learner.}{researchtool}
Logit lens and activation patching were built for questions a researcher cannot yet answer, and their conclusions rest on aggregation across many prompts.
Single-prompt runs do occur in research practice, but primarily to build intuition for the experiments that follow.
In an educational setting, one prompt may be both the start and the end of a learner's exploration. Formative participants could not establish the generality of what they produced (\S\ref{sec:formative}).
A patch that does not move the output is informative to a researcher, and directs the next revision of the setup.
For a learner it is indistinguishable from a badly constructed prompt and from a broken tool.
 
\challengehead{Mechanistic results do not carry their own justification.}{justification}
Each mechanistic technique has a bounded conclusion, and the bound is not visible in the technique's output.
Logit lens is observational and lossy (\S\ref{sec:LogitLens}), and patching yields a local causal claim about a single activation in a single contrastive context.
Expert practice supplies these qualifications automatically. The interface does not.
Both end-users and ML practitioners apply interpretability output past what the methods support \cite{schuff_human_2022, kaur_interpreting_2020}.
A mechanistic result can also unseat a learner's prior model without offering a better one: formative participants reported reduced confidence in their understanding of LLMs even while engaging productively with logit lens (\S\ref{sec:formative}).
 
\challengehead{The model under inspection does not match popular LLM tools.}{modelgap}
Non-experts form mental models from the systems they actually use \cite{zamfirescu-pereira_why_2023}, and a deployed assistant shows only its finished reply, concealing most of the mechanism that inspection reveals.
The mismatch is a seam in the sense of \citet{ehsan_seamful_2024} (\S\ref{subsec:seamful}), and formative participants struggled to reconcile a prediction at every position with the one-reply-per-turn behavior they knew (\S\ref{sec:formative}).
Open base models lack the instruction tuning, chat templates, memory, and tool access a regular user of ChatGPT or Claude expects. The tools that put a model in front of a learner have so far traded scale against learner control (\S\ref{subsec:tools}).
Serving inspectable models through remote infrastructure \cite{fiotto-kaufman_nnsight_2024} closes the scale gap but not the behavioral one.

\section{Design Goals}
\label{sec:designgoals}
From the five design challenges we identified (\S\ref{sec:challenges}), we derive five design goals (\ref{goal:logitlens}--\ref{goal:ndif}).
 
\goalhead{Render the residual stream as tokens in model coordinates.}{logitlens}
A transformer diagram and a logit lens readout are each half of a legible picture.
A diagram of the residual stream and transformer blocks gives structure without content: the learner sees where an activation sits but nothing about what it holds without being able to reason about vectors.
Logit lens gives contents without structure: internal state appears as ranked words a learner can read rather than a scalar to decode, on axes \ref{ch:unreadable} shows non-experts could not map back to the model.
We therefore join the two, displaying the lens on the stream, so each readout is at once a legible approximation of an activation and a located point in the model.
That join supplies the two criteria \ref{ch:unreadable} finds wanting: the state appears in the model's own vocabulary, and one display carries structure and contents together.
It also supplies the third, \emph{flexible manipulability}: the coordinates at which a learner reads an activation are those at which they change it, so no separate representation stands between inspection and intervention.
 
\goalhead{Guided experiments precede independent use.}{scaffold}
A learner can operate an intervention without being able to pose a question it answers or read what its result establishes (\ref{ch:workflow}).
Unguided exploration in simulation environments reliably underperforms guided variants \cite{de_jong_scientific_1998}, and this held for language models specifically (\S\ref{subsubsec:surfacelevel}).
We therefore supply the hypothesis and the interpretation before asking a learner to run an intervention, so a first experiment returns a result whose meaning is already given.
Activities move from fully worked examples to partially specified problems to independent construction \cite{renkl_structuring_2003}.
Within an activity, hints are requested rather than imposed, and escalate through the three degrees of guidance established for visual analysis \cite{ceneda_characterizing_2017}: orienting the learner toward what to look at and what has been explored, directing them to candidate prompts and data locations, and prescribing a working intervention loaded on request.
 
\goalhead{Interpretable-by-construction exploration.}{interp}
Exploration must stay open to learner-chosen prompts while ensuring results can support a conclusion (\ref{ch:researchtool}): we commit to both rather than trading one against the other.
Every prompt on the guided path is verified in advance to produce an interpretable trace, so a learner's first encounter with each technique is one where it clearly works.
Onboarding guidance recommends explaining a tool in terms of the example in front of the learner, at the acknowledged risk that a concrete grounding narrows transfer to new cases \cite{stoiber_design_2023}. Own-prompt entry therefore remains available at every step rather than only after the guided path.
Where \ref{goal:scaffold} supplies the structure of an experiment, this goal governs what the experiment returns.
 
\goalhead{Explanations that carry concept and scope together.}{explain}
Each interface view must be paired with plain language explaining what the technique shows and what it does not, in the same place at the same time (\ref{ch:justification}).
A caveat delivered separately from the result will not be read alongside it, which makes placement rather than presence the requirement.
The explanations must also supply positive content, because \ref{ch:justification} leaves a learner with doubt and no replacement.
We therefore write the explanations against an articulated account of LLM text generation, so a learner leaves with a better model of the system rather than only doubt about the one they arrived with.
 
\goalhead{Research-scale models with managed expectations.}{ndif}
Inspectable models must be brought closer to the systems users actually use: in scale where possible, in expectation where not (\ref{ch:modelgap}).
Serving models remotely on NDIF \cite{fiotto-kaufman_nnsight_2024} closes the scale gap, on the same open-weight models interpretability researchers study.
The behavioral gap cannot be closed, so we treat expectation-setting as part of the design. Before any intervention, we establish that a prediction is computed at every position, that continuation requires another run, and that the model draws on nothing but its weights and the prompt (\S\ref{sec:transformers}).

\section{PatchLens Interface}
\label{sec:patchlens}
PatchLens is a browser-based interface for running logit lens and activation patching on open-weight models hosted by NDIF \cite{fiotto-kaufman_nnsight_2024}.
One gesture on the lens grid performs every intervention PatchLens supports: a click reads a cell out (\S\ref{subsec:locating}), and a drag from one prompt onto another patches it (\S\ref{subsec:patching}).
Guided activities (\S\ref{subsec:tutorial}) and in-place explanations (\S\ref{subsec:explain}) support use without an instructor.

\subsection{The Lens Grid}
\label{subsec:lensgrid}

\begin{figure}[t]
  \centering
  \begin{subfigure}[t]{0.70\linewidth}
    \includegraphics[width=\linewidth]{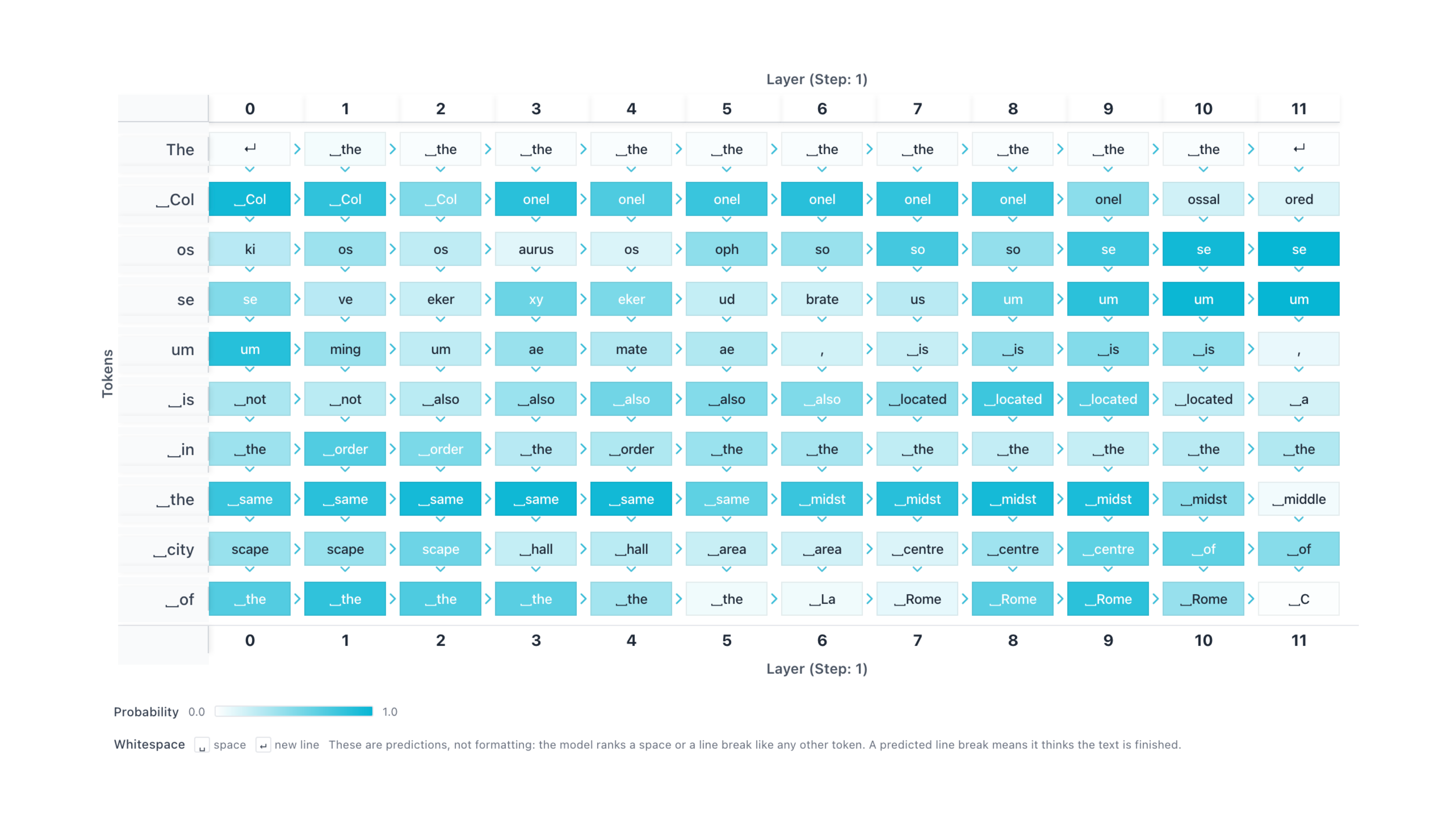}
    \caption{Every layer at once, with arrows tracing the residual stream. The grid fits 8--12 layers on screen depending on window size, and the last cell of the last row is the model's next-token prediction.}
    \Description{A grid of text cells. Rows are labeled with the input tokens of the prompt "The Colosseum is in the city of", one token per row, with "Colosseum" split across three rows. Columns are labeled 0 through 11, one per layer of GPT-2. Each cell holds a single token in a light blue box whose shading darkens with higher probability, and a small downward arrow below each cell links it to the next layer, tracing the residual stream left to right. Early columns hold unrelated tokens such as "the", "in", and fragments; later columns converge, and the final cell of the bottom row holds the model's next-token prediction. A probability legend runs along the bottom from 0.0 in white to 1.0 in dark blue.}
    \label{fig:expanded}
  \end{subfigure}
  \hfill
  \begin{subfigure}[t]{0.26\linewidth}
    \includegraphics[width=\linewidth, trim={534 600 1150 114}, clip]{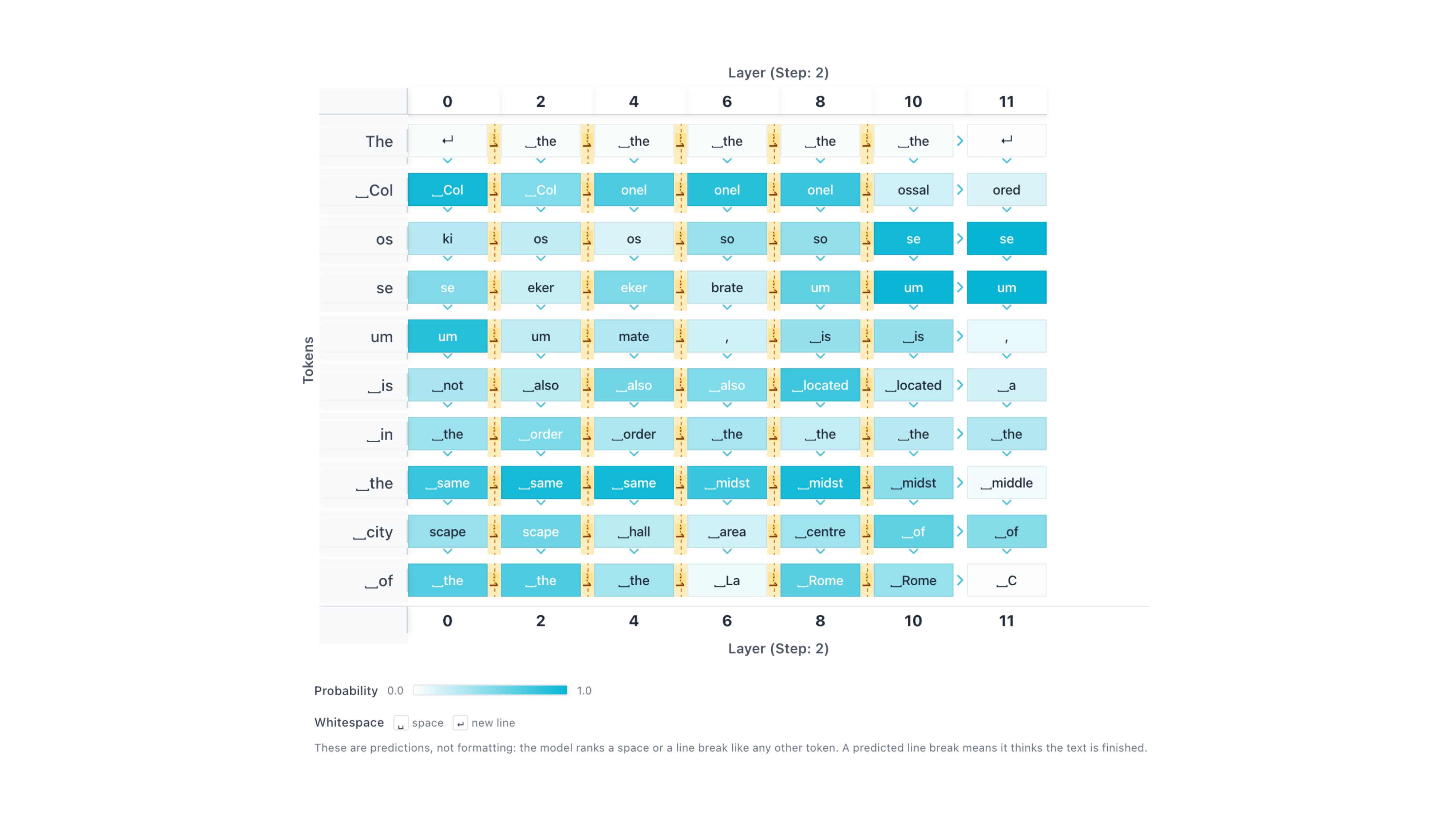}
    \caption{The same prompt with a layer step of 2. Layers 1 and 3 fold into the amber band.}
    \Description{A narrow two-column crop of the same grid, showing only the columns labeled 0 and 2. Between them runs a vertical amber band standing in for layers 1 and 3, which are folded away at this layer step. The cells shown hold the tokens for the same input positions as in the full grid, so the reader sees layer 0 and layer 2 side by side with the intervening layers collapsed.}
    \label{fig:collapsed}
  \end{subfigure}
  \caption{The lens grid on ``The Colosseum is in the city of'', on GPT-2, small enough that every layer fits the page; most models PatchLens serves are not, which is what the layer step in (b) addresses. Darker shading is higher probability, and a word split into several tokens occupies several rows.}
  \label{fig:lensgrid}
\end{figure}

The lens grid (Fig.~\ref{fig:lensgrid}) renders the residual stream as tokens in model coordinates: each row is an input token position and each column a model layer (\ref{goal:logitlens}).
One prompt produces one grid, and the workspace holds up to two prompts, plus a third grid for a patched run (\S\ref{subsec:patching}).
Previously run prompts remain available in a history panel (Fig.~\ref{fig:teaser}.A).

Cell contents follow the conventional logit lens encoding (\S\ref{sec:LogitLens}), the top-ranked token with one scalar rendered as shading, and a click reveals the ranked list (\S\ref{subsec:locating}).
Arrows between cells trace the residual stream forward, a relationship between activations that a plot cannot show.
Larger models have far more layers than fit on screen, Llama-3.1-405B has 126, so layer bands are collapsible and the layer step can be customized (Fig.~\ref{fig:collapsed}).

\subsection{Reading a Cell in Context}
\label{subsec:locating}

\begin{figure}[t]
  \centering
  \includegraphics[width=\linewidth]{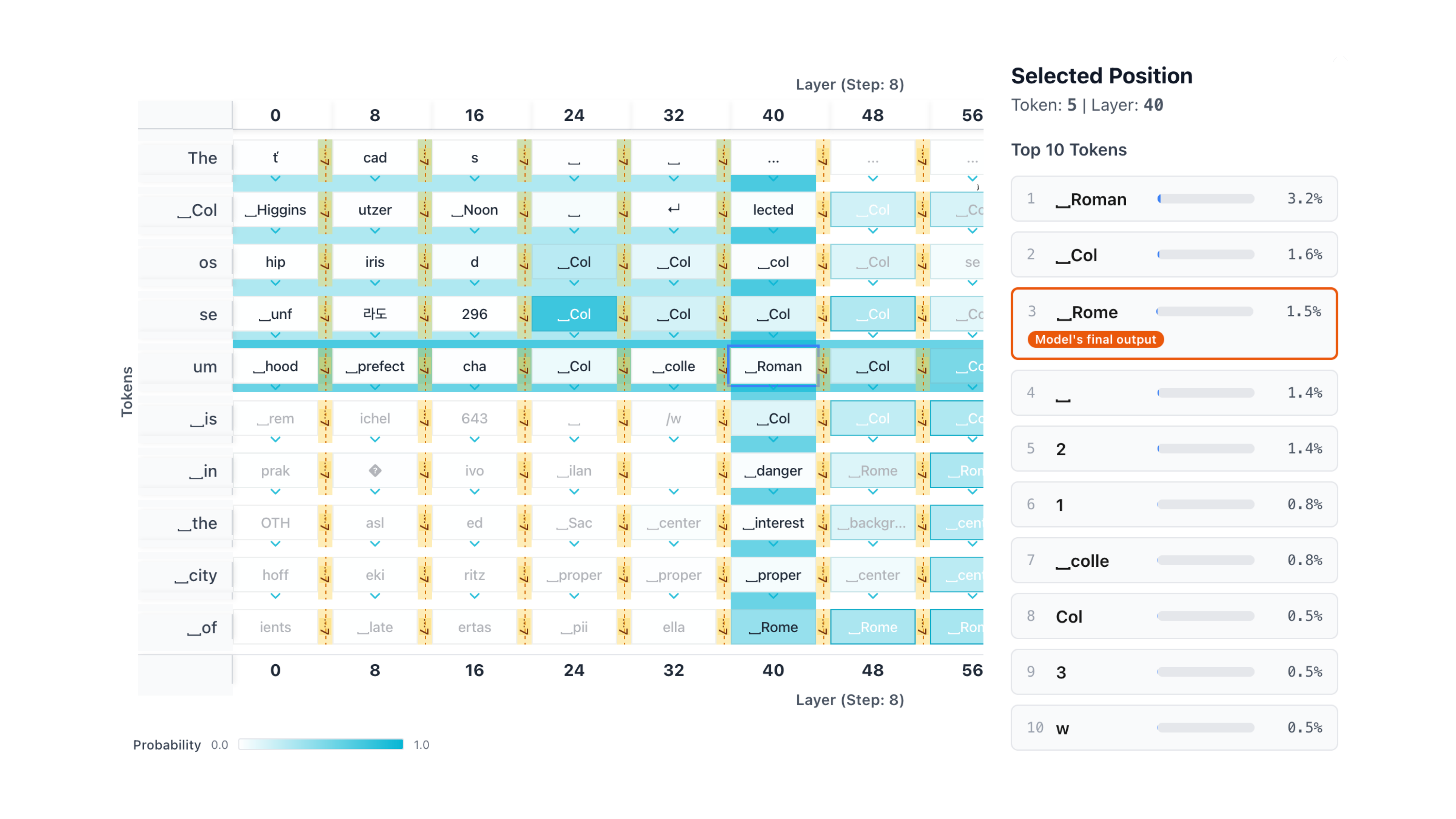}
  \caption{Layer 40 at input position 5 (``um'') selected, for ``The Colosseum is in the city of'' on Llama-3.1-70B-Instruct. The panel lists the ten highest-ranked tokens at that position. Orange marks the model's output after the whole prompt, here ``Rome''; green would mark the final-layer prediction for the token after ``um'', which falls outside the top ten. Behind the panel the grid is re-rendered around the selection (\S\ref{subsec:locating}).}
  \Description{A lens grid for "The Colosseum is in the city of" on Llama-3.1-70B-Instruct, with columns labeled 0, 8, 16, 24, 32, 40, 48, and 56 at a layer step of 8, beside a panel titled "Selected Position". The panel reports token 5, layer 40, and lists the ten highest-ranked tokens at that position with their probabilities, from "_Roman" at 3.2 percent down to "w" at 0.5 percent. The third entry, "_Rome" at 1.5 percent, is boxed in orange and tagged "Model's final output". Behind the panel the grid is re-rendered around the selected cell: its row and its column are highlighted, a shaded region extends up and to the left of the cell covering earlier layers at this and earlier positions, and all remaining cells are dimmed. A probability legend runs along the bottom from 0.0 to 1.0.}
  \label{fig:position}
\end{figure}

For logit lens the gesture is a click on a cell.
The click opens a panel of ranked readouts for that position and re-renders the grid around the selection (Fig.~\ref{fig:position}).

The emphasized region behind the cell is the causal mask (\S\ref{sec:transformers}), which restricts each activation to earlier positions and layers, and the interface calls it the \textit{cone}.
It reaches left and up, never forward or down, so with the forward arrows of \S\ref{subsec:lensgrid} the grid shows both where a cell's information can go and what was available to it (\ref{goal:logitlens}).
We also adapted the crosshair convention of spreadsheets and matrix views, where selecting a cell lights its row and column, an idiom many non-experts arrive able to read (\ref{goal:logitlens}).

Cells outside these regions are dimmed rather than hidden, so the selection can be read against the whole grid.
The \textit{row} shows one input position's predicted token refinement across layers, and the \textit{column} is one layer across every position, each with its own readout.
The column is where a learner sees a prediction take shape at every input position and not only at the last (\S\ref{sec:transformers}, \ref{ch:modelgap}).

\subsection{Patching as Direct Manipulation}
\label{subsec:patching}

\begin{figure}[p]
  \centering
  \begin{subfigure}[t]{\linewidth}
    \includegraphics[width=0.88\linewidth, trim={194 163 191 72}, clip]{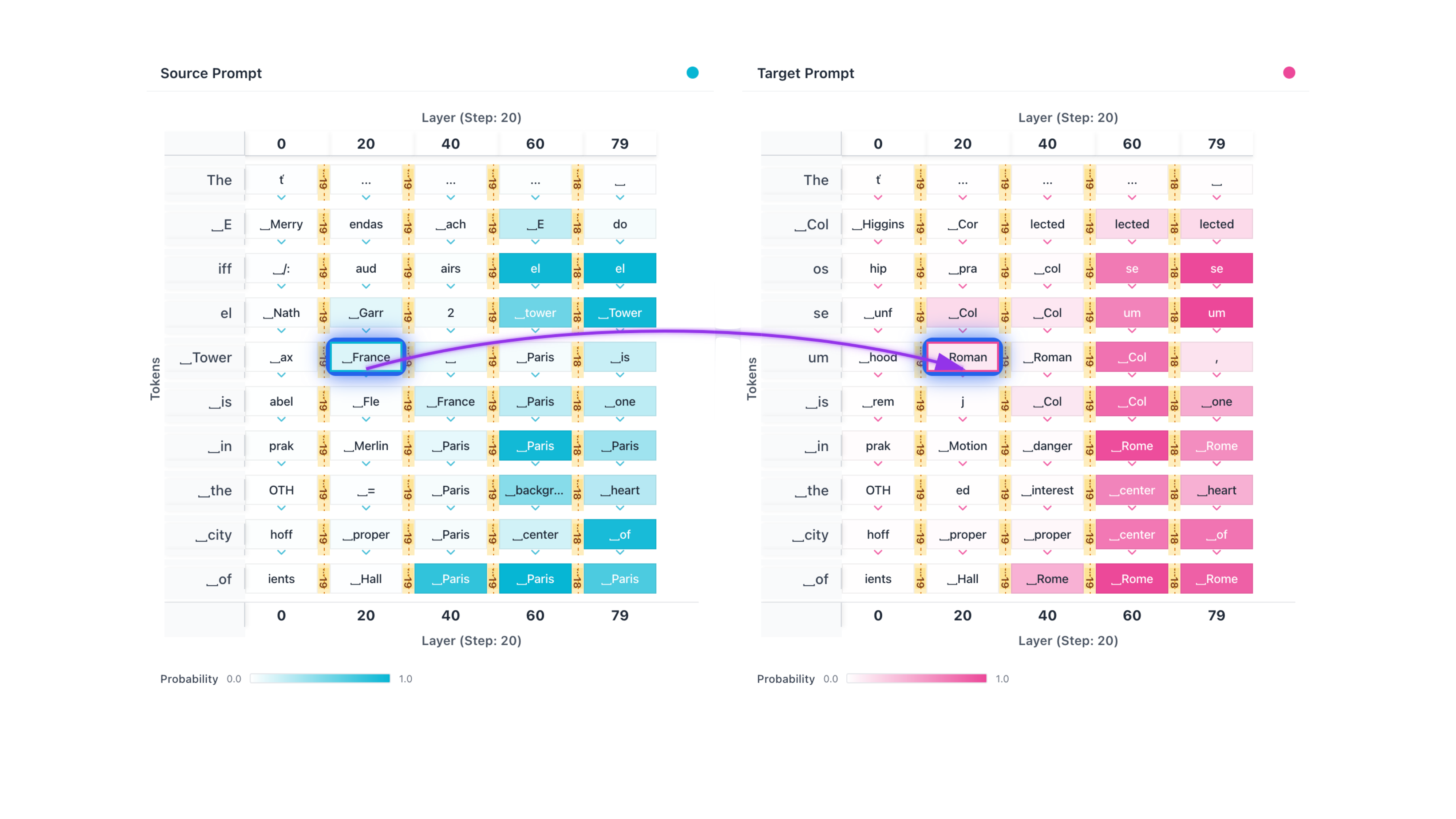}
    \caption{After the drag-and-drop patching action. The purple arrow marks the cells that have been patched.}
    \Description{Two lens grids side by side, each with columns labeled 0, 20, 40, 60, and 79 at a layer step of 20. The left grid, headed "Source Prompt", is shaded blue and holds the tokens of "The Eiffel Tower is in the city of"; the right grid, headed "Target Prompt", is shaded pink and holds "The Colosseum is in the city of". A cell in the source grid at layer 20, on the row for the token "Tower", holds "_France" and is outlined in blue. A curved purple arrow runs from that cell to a cell in the target grid at the same layer, on the row for the token "um", which holds "_Roman" and is outlined in purple, marking the activation that has been dragged from the source into the target.}
    \label{fig:patchaction}
  \end{subfigure}

  \vspace{0.5em}
  \begin{subfigure}[t]{\linewidth}
    \includegraphics[width=0.88\linewidth, trim={249 163 250 79}, clip]{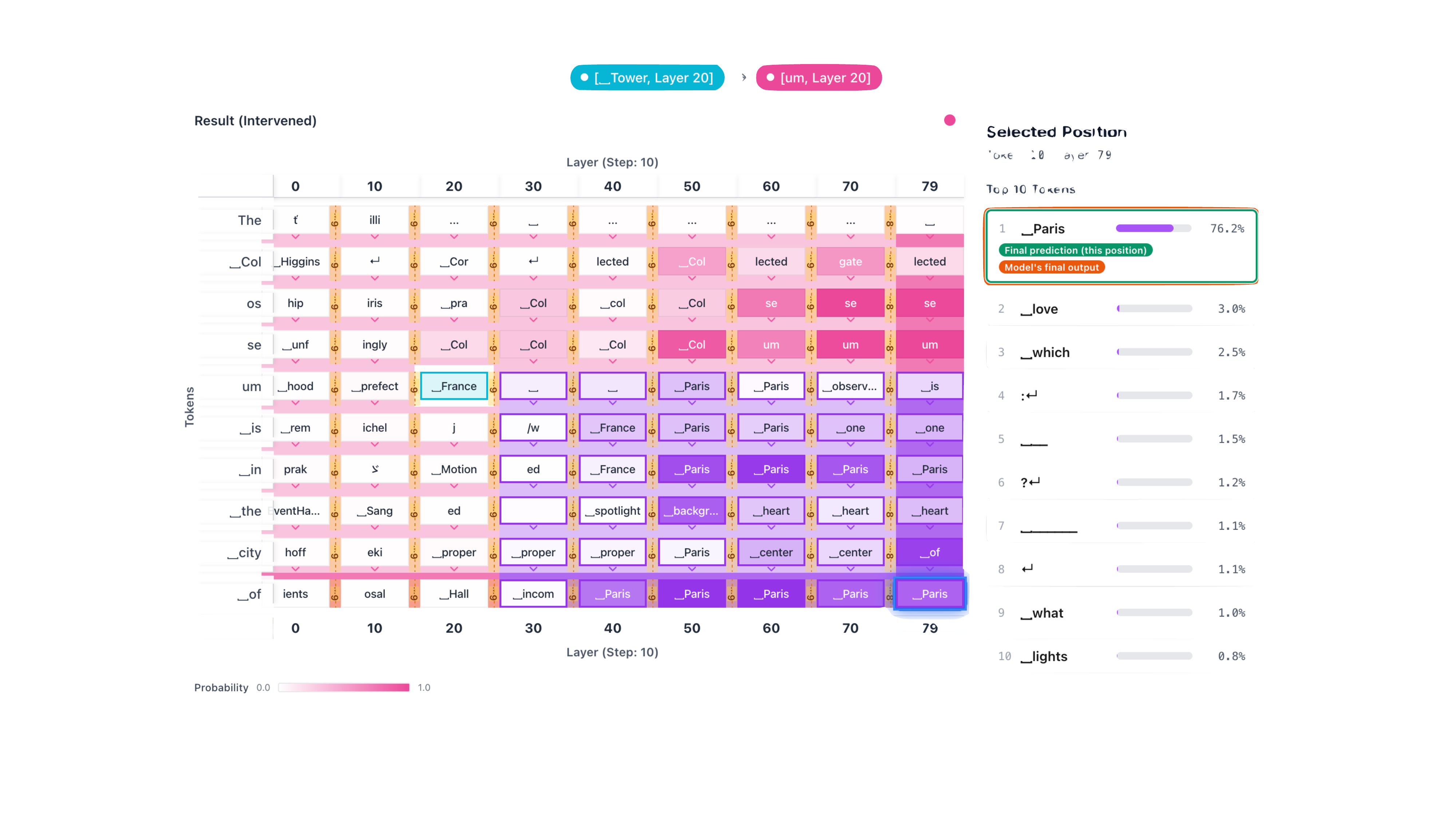}
    \caption{The patched run. Blue is the inserted activation, the source prompt's color; pink marks cells the patch did not reach and purple those downstream of it whose values changed. The final token is now ``Paris'', the source's output.}
    \Description{A third lens grid headed "Result (Intervened)", with columns labeled 0 through 79 at a layer step of 10, beside a "Selected Position" panel. In the grid, one cell at layer 20 on the row for the token "um" is outlined in blue, marking the inserted activation carried over from the source prompt. Cells downstream of it are shaded purple, and cells the patch did not reach remain pink. The final cell of the bottom row, at layer 79, holds "_Paris" and is outlined in blue. The panel lists the ten highest-ranked tokens at that final position: "_Paris" at 76.2 percent, tagged both "Final prediction (this position)" and "Model's final output", followed by "_love" at 3.0 percent, "_which" at 2.5 percent, and seven further tokens at or below 1.7 percent.}
    \label{fig:patchresult}
  \end{subfigure}
  \caption{A patching experiment on Llama-3.1-70B-Instruct. Source: ``The Eiffel Tower is in the city of''; target: ``The Colosse\textbf{um} is in the city of''. The activation at input position 5 and layer 20 is patched.}
  \label{fig:patching}
\end{figure}

For the contrast to be interpretable the source and target must differ minimally, as in the running example of ``The Eiffel Tower is in the city of'' and ``The Colosseum is in the city of''.
A drag from a cell of the source grid onto the target grid at the same layer patches that activation into the target's forward pass (Fig.~\ref{fig:patchaction}), so the learner performs the patch rather than specifying it.
A third grid then appears below the two showing the patched run (Fig.~\ref{fig:patchresult}).
The learner drops a cell whose contents they have already read onto a cell whose contents they have already read, and the result is displayed in the same frame, at the same coordinates, in the same vocabulary (\ref{goal:logitlens}).

The learner establishes that the tokens they read in that cell were information the target's output depended on, not merely something present in the model at that point: the difference between what a model represents and what it uses (\S\ref{sec:background}), produced by the learner's own intervention rather than asserted by the interface.
PatchLens deliberately patches a single layer and position in a single prompt pair, which bounds that claim to one site.
It does not establish that the information is stored only there, or that the same site matters for a different pair of prompts, which only the sweep a research patching experiment performs (\S\ref{sec:ActPatch}) would support.

\subsection{Guided Activities}
\label{subsec:tutorial}

\begin{figure}[htbp]
    \centering
    \includegraphics[width=\linewidth, trim={408 168 582 192}, clip]{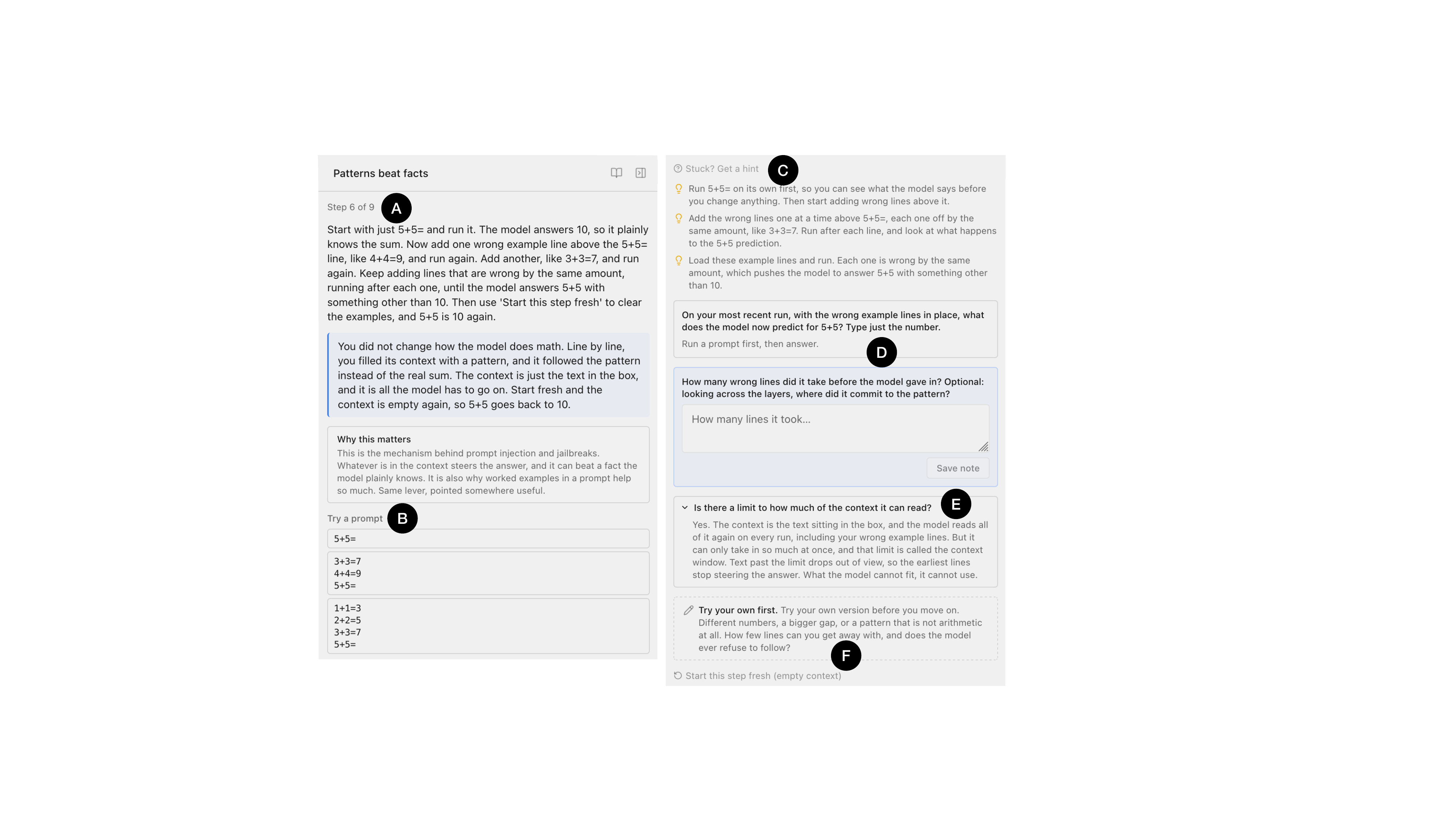}
    \caption{A unit of a guided activity, as the learner sees it, here teaching multi-shot prompting. (A) The task, the concept it demonstrates, and why that concept matters outside the tool. (B) Author-supplied prompts to load and run. (C) Hints, escalating from orienting to a working configuration; none are shown until requested. (D) A check on the model's output and a free-text note; neither gates progress. (E) A unit FAQ. (F) A suggestion for the learner's own prompt, and a control clearing the unit.}
    \Description{A two-column interface panel for a guided activity unit titled "Patterns beat facts", marked step 6 of 9. The left column gives the task in prose: run the prompt "5+5=" on its own, then add lines above it that are wrong by a consistent amount, such as "3+3=7" and "4+4=9", and keep adding until the model answers 5+5 with something other than 10. Below the task sits an explanation of why the result happens, a "Why this matters" box connecting it to prompt injection and worked examples, and a "Try a prompt" area holding three author-supplied prompts of increasing length. The right column holds a collapsed "Stuck? Get a hint" section listing three hints in escalating order, a free-response check asking what the model now predicts for 5+5, an optional note field with a "Save note" button, an expandable unit FAQ headed "Is there a limit to how much of the context it can read?", a "Try your own first" suggestion, and a control labeled "Start this step fresh (empty context)".}
    \label{fig:tutorialprog}
\end{figure}

PatchLens works as a standalone tool or with an instructor-authored guided activity.
Guided activities are sequences of units, each specifying a task, the concept it demonstrates, and a prompt of the learner's own to try (Fig.~\ref{fig:tutorialprog}).
An author can specify the prompt(s) a unit loads, the layer and position it directs attention to, and the prompt pair a patching unit uses, so every experiment configuration a learner meets on the guided path has been verified in advance (\ref{goal:interp}).
Each interaction can also be given hints, which an author escalates from orienting the learner toward what to look at, to directing them to specific rows, columns, and controls, to prescribing a working configuration the interface loads on request (\ref{goal:scaffold}) \cite{ceneda_characterizing_2017}.
Authors work in the application, in data rather than code, so guided material can be added or replaced without rebuilding the tool.

\subsection{In-Place Explanations}
\label{subsec:explain}

\begin{figure}[t]
    \centering
    \includegraphics[width=\linewidth]{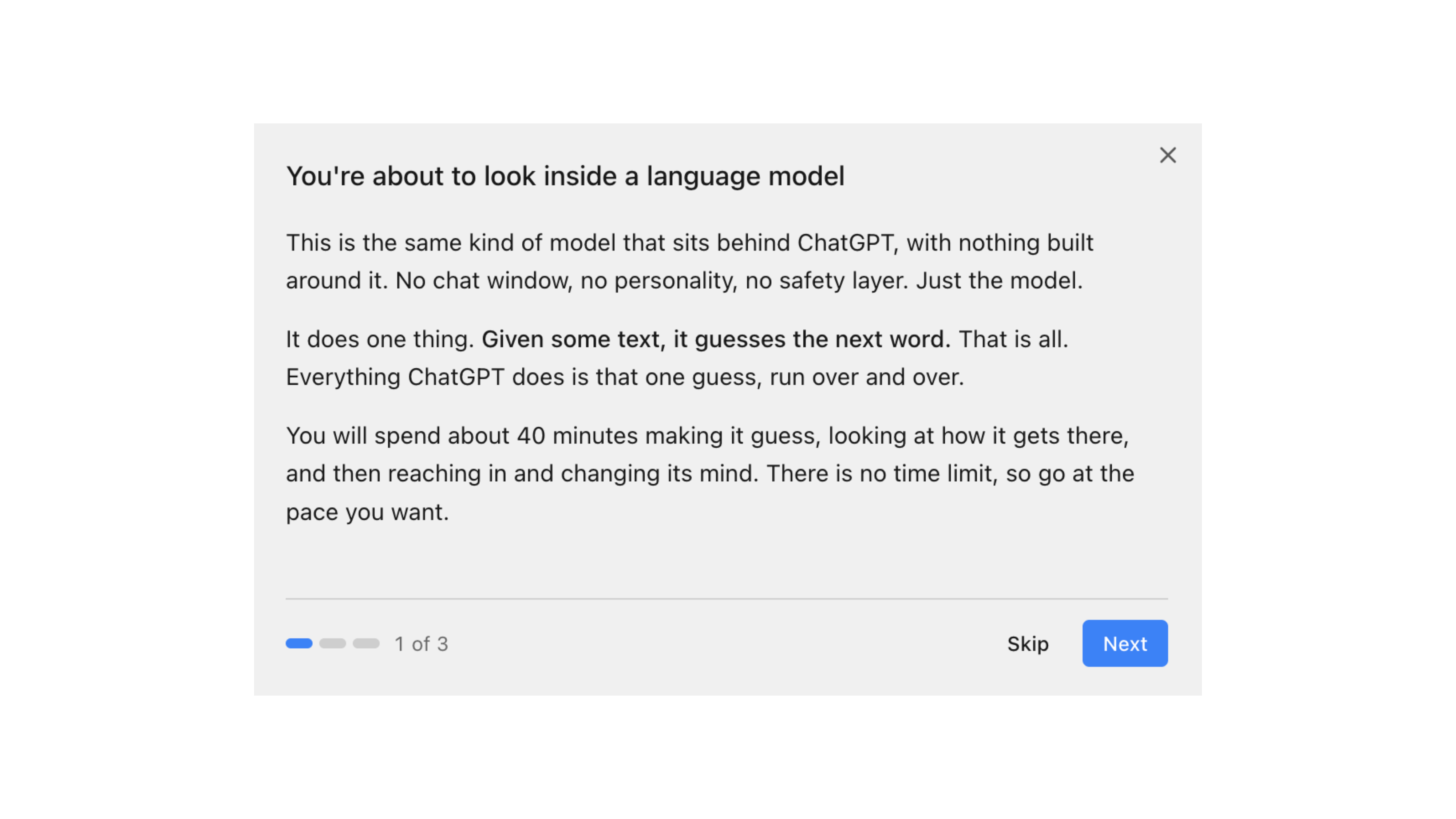}
    \caption{The first of three welcome slides, shown before any intervention. It names the gap between this model and the assistants a learner has used, and states the one operation the interface is built around.}
    \Description{A modal dialogue headed "You're about to look inside a language model", shown over a dimmed interface. Three short paragraphs state that this is the same kind of model that sits behind ChatGPT with nothing built around it, no chat window, personality, or safety layer; that it does one thing, guessing the next word from some text, and that everything a chat assistant does is that guess run over and over; and that the learner will spend about 40 minutes making it guess, looking at how it gets there, and then reaching in and changing its mind, with no time limit. A progress indicator reads "1 of 3", beside a "Skip" link and a "Next" button.}
    \label{fig:expectations}
\end{figure}

Whereas an activity varies by author, a second layer of explanation is fixed and reaches every learner (\ref{goal:explain}).
Because expectations formed by assistants such as Claude or ChatGPT do not transfer to a model under inspection (\ref{goal:ndif}), a welcome sequence names that gap before any intervention is run (Fig.~\ref{fig:expectations}).
Three units of the activity used in our study (\S\ref{sec:methods}) then take up the three facts of \S\ref{sec:transformers} that \ref{goal:ndif} names.

\begin{table}[t]
  \caption{The six explanation surfaces in PatchLens. Surfaces marked $\dagger$
  are authored per activity and vary with it; the rest are fixed and reach every
  learner. Escalation refers to the three degrees of guidance of
  \citet{ceneda_characterizing_2017} (\ref{goal:scaffold}).}
  \Description{A table with six rows, one per explanation surface, and three
  columns: the surface, its scope, and what it carries. Read as surface, scope,
  carries: Welcome sequence, before any run, the gap between the model and the
  assistant. Orientation walkthroughs, per interface region, operating and
  reading a view. Grid reference, any step, what a cell holds. Glossary, any
  step, the interface's own terms. Unit hints, per interaction on request,
  escalating guidance. Unit FAQ, per unit, the questions that unit raises. A
  dagger follows the surface name on the unit hints and unit FAQ rows, marking
  the two surfaces that are authored per activity; the other four are fixed and
  reach every learner. Every entry in the third column also carries a reference
  to the design goal that surface serves.}
  \label{tab:surfaces}
  \begin{tabular}{@{}lll@{}}
    \toprule
    Surface & Scope & Carries \\
    \midrule
    Welcome sequence & Before any run & The model/assistant gap (\ref{goal:ndif}) \\
    Orientation walkthroughs & Per interface region & Operating and reading a view (\ref{goal:explain}) \\
    Grid reference & Any step & What a cell holds (\ref{goal:explain}) \\
    Glossary & Any step & The interface's own terms (\ref{goal:explain}) \\
    Unit hints$\dagger$ & Per interaction, on request & Escalating guidance (\ref{goal:scaffold}) \\
    Unit FAQ$\dagger$ & Per unit & Questions that unit raises (\ref{goal:explain}) \\
    \bottomrule
  \end{tabular}
\end{table}

\subsection{Implementation}
\label{subsec:implementation}

PatchLens runs in the browser against models served remotely on NDIF \cite{fiotto-kaufman_nnsight_2024} (\ref{goal:ndif}).
Any model NDIF serves can be selected from a menu that groups base and chat variants, ranging from GPT-2 to Llama-3.1-405B-Instruct.
We use Llama-3.1-70B-Instruct throughout this paper, and GPT-2 in Fig.~\ref{fig:lensgrid}.

PatchLens derives its top tokens and probabilities from raw logit lens rather than from a fitted lens.
The tuned lens (\S\ref{sec:LogitLens}), Future Lens \cite{pal_future_2023}, and the Jacobian lens \cite{gurnee_verbalizable_2026} must each be fitted per model, which is technical overhead and a constraint on how many models a tool can host, where raw projection works on whatever NDIF serves (\ref{goal:ndif}).
Logit lens is also the model's own operation, applying the same unembedding matrix that produces the model's output, where a fitted lens interposes a step the model does not perform and the interface would have to explain before its readouts could be read.
The cost is fidelity, and we state the limits of raw projection where the readouts appear (\ref{goal:explain}).
PatchLens is publicly available; we omit the link to preserve anonymity during review.

\section{Design Iteration}
\label{sec:iteration}
The design in \S\ref{sec:patchlens} is the product of repeated revision: we tested working versions informally with colleagues and non-expert volunteers throughout development, and ran one structured pilot of the first complete version.

We evaluated the first version of PatchLens with 10 philosophy graduate students from a later cohort of the AI summer school that hosted the formative participants (\S\ref{sec:formative}), all reporting little or no prior AI background.
They had two hours with PatchLens and a guided worksheet, after a short instructor-led introduction to both techniques.
They then completed the System Usability Scale (SUS) and open-ended questions on what they found helpful, confusing, and would improve.
Nine of the 10 completed the SUS, giving a mean of 51.9 (range 32.5 to 75), below the 68-point benchmark for acceptable usability.
At this sample size the score locates problems rather than measuring usability \cite{sauro_chapter_2016}.
We read the open-ended responses for areas to improve without coding them formally.

The changes below are all reflected in the version evaluated in \S\ref{sec:methods}, except the two noted at the end.
Participants found the interface manageable but could not navigate it without external guidance, and asked for more detailed worksheet steps.
Because PatchLens must be usable outside a workshop, we moved that guidance into the tool as an in-application walkthrough introducing each control in context (\ref{goal:scaffold}).
Both the layer step selector and the orange fill marking the final predicted token were singled out as opaque, so we added an inline tooltip to the first and removed the second (\ref{goal:explain}).
Several respondents had not retained the distinction between the two techniques from the instructor-led introduction, so we expanded the in-tool tutorial and linked concept explanations from each view (\ref{goal:explain}).
Two lost confidence in the tool after prompts with trailing whitespace produced misleading output, a correctness bug rather than a design fault.
We normalized whitespace during preprocessing.

We declined two requests. The interface struck two participants as crowded.
We emphasized the final prediction cell, which participants overlooked (\ref{goal:logitlens}), but kept the grid dense, because density is what shows the progression across layers at once.
We also declined patching at several locations, which would cost this iteration's conceptual clarity (\ref{ch:researchtool}).

\section{Evaluation Methods}
\label{sec:methods}
The evaluation compares the two tools on what participants learned, what they chose to run, and how they experienced the session:
\begin{enumerate}
    \item[\textbf{RQ1}] How does understanding of LLM text generation compare after using each tool, and how accurately did participants judge their own understanding?
    \item[\textbf{RQ2}] What did participants choose to run when the prompt was their own rather than one supplied by the tool?
    \item[\textbf{RQ3}] How did participants' use of each tool shape their experience of it?
\end{enumerate}

\subsection{Study Design}
\label{subsec:studydesign}

Participants were randomly assigned to a guided learning tutorial for the PatchLens interface or to Transformer Explainer \cite{cho_transformer_2026}, in a between-subjects design chosen to avoid knowledge transfer across conditions. The study was approved by our institutional IRB.
We chose Transformer Explainer as the most widely used interactive resource for learning transformer architecture. No existing tool holds interactivity, novice-friendly explanation, research-scale models, and intervention on internal computation together (\S\ref{subsec:tools}), so no direct comparison on matched content was available.
Both conditions are interactive learning visualizations, which controls for the benefit interactivity itself confers (\S\ref{subsubsec:scaletrade}) and leaves content, including the added mechanistic interpretability techniques, as what differs.

We identified nine learning objectives and thirteen corresponding multiple-choice questions (Table~\ref{tab:los}), constrained to what either condition could teach, and consulted two instructors on the objectives and the instrument.
The knowledge assessment was administered only after tool use, to keep the session within its time budget and to avoid priming participants on the material.
Random assignment supports the between-arm comparison without a pre-test, but the design measures no within-participant learning.

\begin{table}[t]
  \caption{Learning objectives (LOs), the knowledge-assessment items that measure
  each, and the matched self-rated understanding items. Knowledge items are keyed
  multiple-choice (Table~\ref{tab:instrument}); self-rating items are five-point
  Likert statements of the form ``I could explain \ldots'' administered after the
  knowledge block.}
\Description{A four-column table with one row for each of nine learning
  objectives. The columns are the objective label, its description, the knowledge
  item or items that assess it, and the matched self-rating item. Read as
  objective, description, knowledge items, self-rating item: LO1, recognize and
  identify large language models and generative AI as distinct from other kinds of
  AI, K11, S1. LO2, understand a large language model as a next-token predictor,
  K1 and K2, S2. LO3, understand that text is produced by running the model many
  times, one token at a time, K3, S3. LO4, identify what is and is not in the
  context window, including that the application rather than the model assembles
  it, K4, K5, and K6, S4. LO5, understand how examples in few-shot prompting add
  context, K7, S5. LO6, recognize why large language models have token and context
  limits, K12, S6. LO7, recognize that incorrect output is a property of the
  generation mechanism rather than a bug, K8, S7. LO8, distinguish the model is confident from the model is correct, K9 and K13, S8. LO9, distinguish the model
  produced X from the model retrieved X, K10, S9. Every objective maps to exactly
  one self-rating item, S1 through S9. The thirteen knowledge items are
  distributed unevenly: LO4 carries three items, LO2 and LO8 carry two each, and
  the remaining six objectives carry one each.}
  \label{tab:los}
  \small
  \begin{tabular}{@{}llcc@{}}
    \toprule
    \textbf{LO} & \textbf{Description} & \textbf{Knowledge} & \textbf{Self-rating} \\
    \midrule
    LO1 & Recognize and identify LLMs and generative AI, as distinct from other kinds of AI & K11 & S1 \\
    LO2 & Understand an LLM as a `next-token predictor' & K1, K2 & S2 \\
    LO3 & Understand that text is produced by running the model many times, one token at a time & K3 & S3 \\
    LO4 & Identify what is and is not in the `context window', including that the & K4, K5, K6 & S4 \\
        & \quad application, not the model, assembles it & & \\
    LO5 & Understand how examples (few-shot prompting) add context & K7 & S5 \\
    LO6 & Recognize why LLMs have token and context limits & K12 & S6 \\
    LO7 & Recognize that incorrect output is a property of the generation mechanism, not a bug & K8 & S7 \\
    LO8 & Distinguish `the model is confident' from `the model is correct' & K9, K13 & S8 \\
    LO9 & Distinguish `the model produced X' from `the model retrieved X' & K10 & S9 \\
    \bottomrule
  \end{tabular}
\end{table}

A post-hoc review of each item's stem and response options against both tools' content found one item dependent on vocabulary specific to one condition.
K7's four response options all refer to the model's weights, a term present in Transformer Explainer but absent from PatchLens's guided path, which teaches the objective K7 measures through examples and patterns.
We retain K7 in the primary estimate and report the comparison without it as a sensitivity analysis, which drops LO5.

\subsection{Participants}
\label{subsec:participants}

We recruited participants from Prolific \cite{noauthor_prolific_2026} for a 60-minute study comprising consent and pre-screening, a pre-survey, guided learning time on one tool, and a post-survey.
To target non-expert learners, we pre-screened for no formal background in AI (no AI/ML research, model-development work, specialized degree, or university-level teaching) and at least moderate interest in how generative AI language tools work.
Participants who completed the study received \$12 and those screened out \$0.14, and the median completion time was 54 minutes.

Analysis is restricted to the 70 approved submissions; data from participants outside that set is excluded from all analysis and from the analysis dataset.
Of the 70, three did not complete the post-survey, leaving 67 (36 PatchLens, 31 Transformer Explainer). 
We excluded six more: four had submitted the pre-survey repeatedly, altering their reported AI background until it satisfied the screener (three PatchLens, one Transformer Explainer), one received two different arm assignments (final assignment to PatchLens), and one completed the post-survey in under five minutes (Transformer Explainer).
No participant was excluded on the basis of a response, though the one time-based exclusion used a duration measured on the survey that contains the knowledge assessment.
Sixty-one participants remained, 32 assigned to PatchLens and 29 to Transformer Explainer.

The median age of the 61 participants was 38 years, and 32 identified as women and 29 as men.
Of the 61 participants, 52 reported using generative AI at least several times a week, and 51 reported not attending any prior AI training, workshop, or professional development.

\subsection{Procedure}
\label{subsec:procedure}

Participants first completed a pre-survey covering demographics, typical AI use, prior AI training or education, AI media exposure, and the eight-item AI self-efficacy scale of \citeauthor{hornberger_multinational_2025}~\cite{hornberger_multinational_2025}.
The pre-survey also administered a single item on mechanistic self-efficacy---agreement with ``I can explain why a language model produced a particular response rather than a different one''---on a five-point Likert scale.

After exploring their assigned tool, participants completed a post-survey in a fixed order.
It opened with the eight-item AI self-efficacy scale and the mechanistic self-efficacy item, both repeated from the pre-survey, then the ten-item System Usability Scale, three engagement items and one comprehensibility item on five-point agreement scales, and one mental demand item.
The thirteen-item knowledge assessment (Appendix~\ref{app:ka}) came next, followed by nine self-rated understanding items, one per learning objective (Table~\ref{tab:los}), and three open-ended items asking what was most helpful, most confusing, and what they would improve.
Participants therefore rated their understanding of each objective having just answered questions on it.

\subsection{Data Analysis}
\label{subsec:analysis}

Throughout, we report estimates with 95\% confidence intervals and effect sizes rather than treating a $p$-value threshold as decisive, and we do not describe any result as showing no difference or as establishing equivalence between the tools.

\subsubsection{Quantitative Analysis}

We drafted the code implementing these quantitative analyses with Claude Code, reviewing both the code and the results carefully.

\paragraph{Knowledge.}
The thirteen knowledge items were scored against a fixed key and analyzed at the item level using a generalized linear mixed model with a binomial logit link.
Assigned interface is a fixed effect, with crossed random intercepts for participant and knowledge item, so the condition effect generalizes over items rather than conditioning on the particular thirteen questions.
We report the condition effect as an odds ratio with a profile-likelihood interval, and participant-level differences in total score with Welch intervals.
We report the minimum detectable effect for this design, computed at 80\% power, and no post-hoc power.
Four sensitivity analyses accompany the primary estimate (\S\ref{sec:results}), including the comparison without K7. The unadjusted estimate on all thirteen items and 61 analyzed participants remains primary throughout.
Accuracy by learning objective is reported descriptively, with no inferential test at the objective or item level, because six of the nine objectives rest on a single item and the intervals at this sample size are too wide for per-objective comparison.

\paragraph{Self-rated understanding and calibration.}
The nine objective-aligned self-rating items were averaged into a composite and compared across arms with a Welch interval.
Per-objective ratings are reported descriptively alongside per-objective accuracy, with the five-point rating rescaled to the accuracy range as $(x-1)/4$ for display.
To assess whether participants could tell which objectives they understood, we fit a mixed logistic model of item correctness on the self-rating for that item's objective, centered on the participant's own mean across the nine and interacted with condition, so the estimate is a within-person slope.

\paragraph{Self-efficacy.}
The eight-item AI self-efficacy scale was administered before and after tool use, and we analyze the post-study score with an analysis of covariance (ANCOVA) including condition and the pre-study mean.
Residual diagnostics triggered a pre-declared rank-based fallback, which we report alongside it.
The single mechanistic self-efficacy item is ordinal. We report its movement within each arm and compare arms with a cumulative-link model on the post-study response, with the pre-study response as a covariate.
Because neither arm was without a tool, within-arm movement common to both is not attributable to either, and we report it descriptively.

\paragraph{Experience measures.}
The four experience measures are on incommensurable scales, so we report a single effect-size family: Cliff's $\delta$ computed from ranks, with bootstrap intervals.

\paragraph{Prompts run.}
We logged the prompts participants ran and count, per participant, the distinct prompts falling outside that arm's example set, once each however many times they were run, in three categories: written from scratch, extended from one of the tool's examples, and extended from a prompt the participant had written themselves.
Both tools demonstrate appending a predicted token to a prompt and re-running it, so no category favors an arm.
PatchLens counts are conservative: because the export records one prompt per run, the second prompt in a patching pair enters the count only when it was also run on its own.

\subsubsection{Qualitative Analysis}

All 61 participants answered all three open-ended items (\S\ref{subsec:procedure}), giving 183 responses: PatchLens respondents averaged 14.0 words per question, Transformer Explainer respondents 16.4.
We treated each response as one analysis unit and allowed more than one code per response.

One author developed an initial codebook from a random 10\% of the responses, then two authors coded 21\% of them together using it, reaching consensus and adding codes as needed.
The two authors then coded the full set independently, reaching Krippendorff's $\alpha = 0.47$ with Measuring Agreement on Set-valued Items (MASI) distance, rising to $0.58$ when codes are grouped into families, which is below conventional thresholds for reliability.
They then met to resolve every difference, so the consensus set underlying every prevalence we report contains no unresolved disagreement.
We therefore report prevalence as descriptive counts on that consensus set, with intervals that overlap between arms on every code, and rest no inferential claim on them.
Where we contrast the arms, the contrast is in what participants chose to write about rather than a measured difference in what they experienced.

Codes were assigned from response content alone, independent of which prompt produced the response, so prevalence is pooled: a code counts once per participant however many of their responses carried it.
Twenty-one thematic codes were applied (eight positive, twelve negative, and one mixed), alongside two administrative codes.
Codes the text does not name are shown in Fig.~\ref{fig:qualcoderesults} for context only. The full codebook, with per-code definitions, prevalence, and agreement, is in Appendix~\ref{app:codebook}.

\section{Results}
\label{sec:results}
Two features of the analyzed sample qualify the comparisons that follow.
Exclusion removed 4 of 36 PatchLens participants against 2 of 31 under Transformer Explainer (\S\ref{subsec:participants}), so the higher-scoring arm is the less heavily filtered one.
The arms were also imbalanced at baseline on three measures: interest in learning about AI was higher under Transformer Explainer (standardized difference $-0.67$), while AI media exposure ($+0.40$) and frequency of generative AI use ($+0.26$) were higher under PatchLens.

\subsection{Knowledge and Self-Assessment (RQ1)}
\label{subsec:rq1-res}

The two arms differ more in how accurately participants judged their own understanding than in what they scored.

\subsubsection{Knowledge}
\label{subsubsec:res-knowledge}

The knowledge comparison does not settle which tool taught more.
PatchLens participants answered 7.44 of 13 knowledge items correctly on average and Transformer Explainer participants 8.66, a difference of $-1.22$ items (95\% CI [$-2.78$, 0.34]).
The item-level model agrees (odds ratio 0.59, 95\% profile-likelihood CI [0.30, 1.16]).
The observed difference falls below the 2.20 items this design could reliably detect, but the interval's lower bound does not, so a detectable Transformer Explainer advantage remains possible.
The instrument bounds it further. With KR-20 $= .756$ ([$.645$, $.827$]), its total-score standard error of measurement is 1.53 items, so the observed difference is 0.80 of one.
The comparison neither establishes a difference in knowledge after tool use nor rules out a Transformer Explainer advantage, though it does exclude a PatchLens advantage larger than about a third of an item.

\begin{figure}
    \centering
    \includegraphics[width=\linewidth]{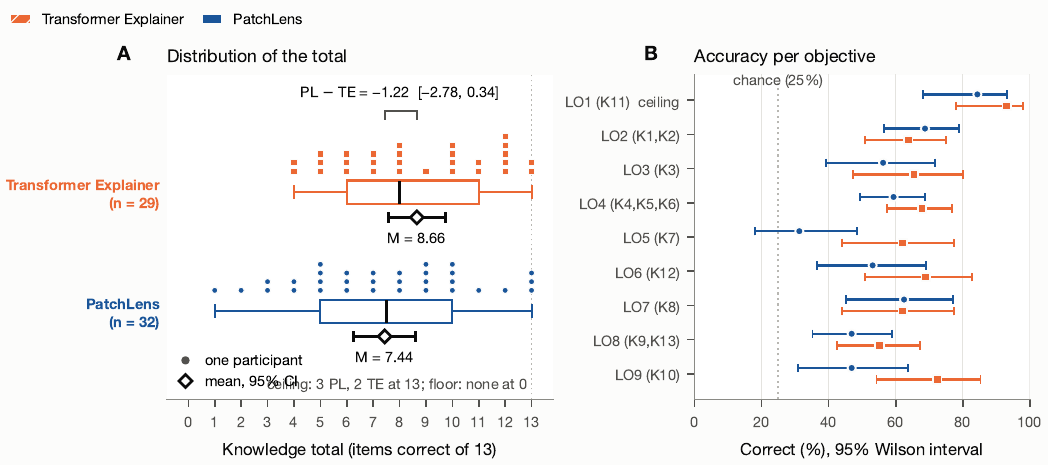}
    \caption{Knowledge totals overlap substantially between arms.
    (A) Each participant's score out of the thirteen items, one dot per participant stacked at their score, over a box giving the median and IQR. Both medians are 8, with IQRs of [5, 10] for PatchLens and [6, 11] for Transformer Explainer, and the diamond below each box is the arm mean with its 95\% interval.
    (B) Accuracy per learning objective, with the 25\% chance line marked.}
    \label{fig:knowledge}
    \Description{Two panels. Panel A is a horizontal dot-and-box plot of knowledge totals, with one row per arm: Transformer Explainer (n = 29) above and PatchLens (n = 32) below. The axis runs from 0 to 13 items correct. Each participant is a dot stacked at their score, over a box giving the median and interquartile range, and a diamond with a horizontal interval below each box marks the arm mean and its 95 percent interval. Both medians sit at 8. The Transformer Explainer mean is 8.66 and the PatchLens mean 7.44, and the two mean intervals overlap substantially. An annotation gives the arm difference as PatchLens minus Transformer Explainer equals minus 1.22, with an interval from minus 2.78 to 0.34. Scores in both arms spread across most of the range, with a few participants at 13 and none at 0. Panel B plots accuracy per learning objective, with nine rows labeled LO1 through LO9 and an axis of percent correct from 0 to 100. Each row carries two point estimates with 95 percent Wilson intervals, one per arm, in the same colors as panel A. A dotted vertical line marks the 25 percent chance level, and every objective sits above it. LO1 is near the ceiling in both arms. The two arms' intervals overlap on every objective, and the Transformer Explainer estimate is higher on seven of the nine.}
\end{figure}

Excluding K7 (\S\ref{subsec:studydesign}) gives $-0.91$ items of 12 ([$-2.36$, 0.54]) and an odds ratio of 0.65 ([0.33, 1.27]), so that single item carries a quarter of the aggregate difference.
Dropping K11 as the item nearest ceiling gives $-1.13$ items ([$-2.65$, $0.39$]) and adjusting for the two imbalanced covariates $-1.28$ ([$-3.07$, $0.51$]).
Restoring the six excluded participants attenuates the estimate furthest, to $-0.54$
([$-2.08$, 0.99]) with an odds ratio of 0.82 ([0.41, 1.61]).
Every point estimate puts Transformer Explainer ahead.

The difference concentrates rather than spreads (Fig.~\ref{fig:knowledge}.B). Point estimates favor Transformer Explainer on seven of the nine, but LO5, LO9, and LO6 carry nearly two thirds of the summed gap. Where PatchLens is nominally higher (LO2 and LO7), the margin is under five percentage points.

\subsubsection{Self-Assessment}
\label{subsubsec:res-selfassess}

Participants rated their own ability to explain each learning objective almost identically across arms, while their measured accuracy differed.
On the nine objective-aligned self-rating items (Cronbach's $\alpha = .95$), PatchLens participants averaged 3.45 of 5 against 3.30 for Transformer Explainer, a difference of $+0.16$ (95\% CI [$-0.33$, $0.65$]), against knowledge item accuracies of 57\% and 67\% respectively.
The self-rating interval includes no difference, and it points in the opposite direction from the accuracy gap.

\begin{figure}
    \centering
    \includegraphics[width=\linewidth]{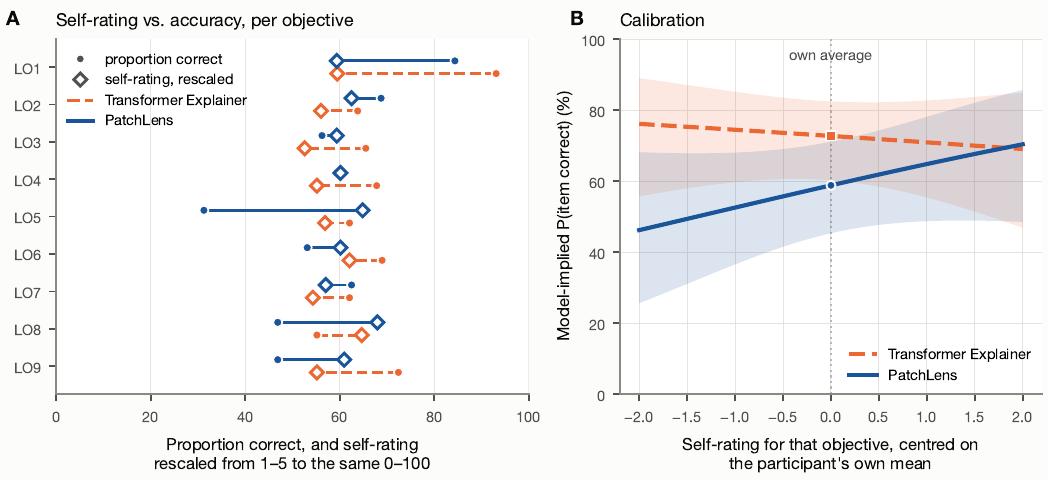}
    \caption{Self-ratings track accuracy unevenly, and in opposite directions in the two arms.
    (A) For each learning objective, the proportion of correct knowledge items (filled circle) and the mean self-rating of ability to explain it (open diamond); the segment length is the discrepancy between them. The five-point rating is rescaled as $(x-1)/4$, a presentational device rather than a claim that a rating of 3 corresponds to 50\% accuracy.
    (B) Within-person calibration is flat for Transformer Explainer ($-0.09$ log-odds per rating point, 95\% CI [$-0.46$, $0.28$]) and positive for PatchLens ($+0.25$, [$-0.12$, $0.62$]), with all three intervals spanning zero. Fifteen participants rated all nine objectives identically and so contribute no within-person variance.}
    \label{fig:selfassessment}
    \Description{Two panels. Panel A has nine rows, LO1 through LO9, and an axis from 0 to 100 on which the proportion of correct knowledge items and the mean self-rating are plotted together, the five-point rating rescaled to the same range. For each objective and each arm, a filled circle marks accuracy and an open diamond marks self-rating, joined by a segment whose length is the discrepancy between them; Transformer Explainer segments are dashed and PatchLens segments solid. The diamonds cluster near 60 across all nine objectives in both arms, while the circles vary widely, so the segments differ mainly in where accuracy falls. Under PatchLens the diamond lies to the right of the circle on six of nine objectives, most conspicuously on LO5, where accuracy is close to 30 percent; under Transformer Explainer it does so on one. Both arms place the circle to the right of the diamond on LO1. Panel B plots within-person calibration: the horizontal axis is a participant's self-rating for an objective centered on their own mean across the nine, running from minus 2 to plus 2, and the vertical axis is the model-implied probability that an item is correct, from 0 to 100 percent. A dotted vertical line at 0 is labeled as the participant's own average. The Transformer Explainer line is dashed and almost flat, running from about 76 percent down to about 70. The PatchLens line is solid and rises from about 46 percent to about 70. Both are surrounded by wide shaded intervals that overlap each other across the whole range.}
\end{figure}

The gap between rated and measured understanding is not uniform across objectives (Fig.~\ref{fig:selfassessment}.A).
Self-rating exceeds accuracy on six of nine objectives under PatchLens but on one under Transformer Explainer.
The mean discrepancy between rescaled self-rating and proportion correct is $+4.8$ percentage points under PatchLens and $-10.5$ under Transformer Explainer.
The objectives on which PatchLens participants scored lowest (LO5, LO8, and LO9, Fig.~\ref{fig:knowledge}.B) have the largest gaps, while self-ratings for the remaining objectives track accuracy more closely in both arms.
The pattern follows accuracy rather than rating. The ratings are nearly flat across the nine objectives (Cronbach's $\alpha = .95$, and fifteen participants used a single value throughout), so the gaps widen wherever accuracy falls.
These are descriptive comparisons on nine single ratings per participant, and no per-objective difference is tested.

Whether participants could tell which objectives they understood is not resolved (Fig.~\ref{fig:selfassessment}.B). The within-person slopes point in opposite directions but all three intervals including the interaction span zero, and only 46 of 61 participants varied their ratings.
The between-person relationship is better resolved. Mean self-rating correlates with knowledge total at $\rho = +0.44$ (95\% CI [$0.06$, $0.75$]) in PatchLens and $\rho = +0.52$ ([$0.21$, $0.74$]) in Transformer Explainer.
Self-ratings therefore track overall knowledge in both arms, though whether they track understanding of particular objectives is not established.

\begin{figure}
    \centering
    \includegraphics[width=\linewidth]{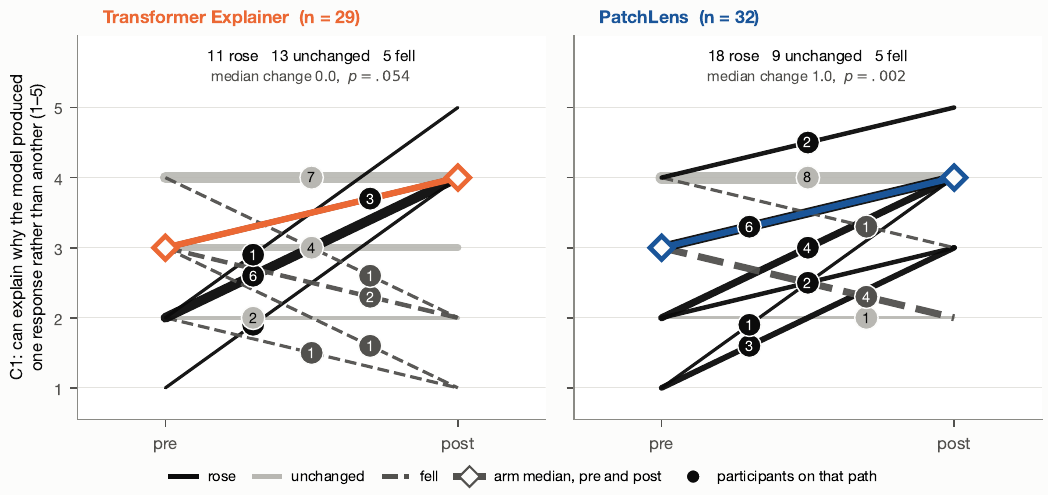}
    \caption{Mechanistic self-efficacy rose for more PatchLens participants than Transformer Explainer participants, with each arm's counts and median change printed in its panel. The item asks participants to rate their agreement with ``I can explain why a language model produced a particular response rather than a different one'' on a 5-point scale (1 = strongly disagree, 5 = strongly agree). One line per distinct response path, its width and label giving the number of participants who followed it, and the diamonds mark each arm's median response at each time.}
    \label{fig:c1-shift}
    \Description{Two side-by-side panels, Transformer Explainer (n = 29) on the left and PatchLens (n = 32) on the right, sharing a vertical axis of agreement with the item from 1 to 5 and a horizontal axis with two positions, pre and post. Each distinct pre-to-post response path is one line, its width and a numeric label giving how many participants followed it; solid black lines rose, grey lines held, and dashed lines fell. A diamond marks each arm's median response at each time, and both arms move from a median of 3 to a median of 4. In the Transformer Explainer panel, 11 participants rose, 13 held, and five fell, for a median individual change of 0.0. In the PatchLens panel, 18 rose, nine held, and five fell, for a median individual change of plus 1.0. No line begins at 5 in either panel. The visible difference between the panels is the weight of the grey held-steady lines, which is heavier on the left.}
\end{figure}

Mechanistic self-efficacy moved further under PatchLens than Transformer Explainer (Fig.~\ref{fig:c1-shift}), though the between-arm contrast does not separate them (cumulative odds ratio 1.56, 95\% CI [$0.57$, $4.30$]). No participant began the study in strong agreement with the statement.
Equal numbers declined in each arm, so the difference lies in how many participants held their answer steady rather than in how many declined.

The eight-item AI self-efficacy scale rose from 3.92 to 4.09 under PatchLens and fell from 3.94 to 3.84 under Transformer Explainer.
Both estimates include no difference. The ANCOVA-adjusted arm difference is $+0.27$ scale points (95\% CI [$-0.05$, $0.59$]), and the rank-based estimate is $+6.41$ rank units ([$-0.84$, $13.66$].

\subsection{Prompts Run (RQ2)}
\label{subsec:rq2-res}

\begin{figure}
    \centering
    \includegraphics[width=\linewidth]{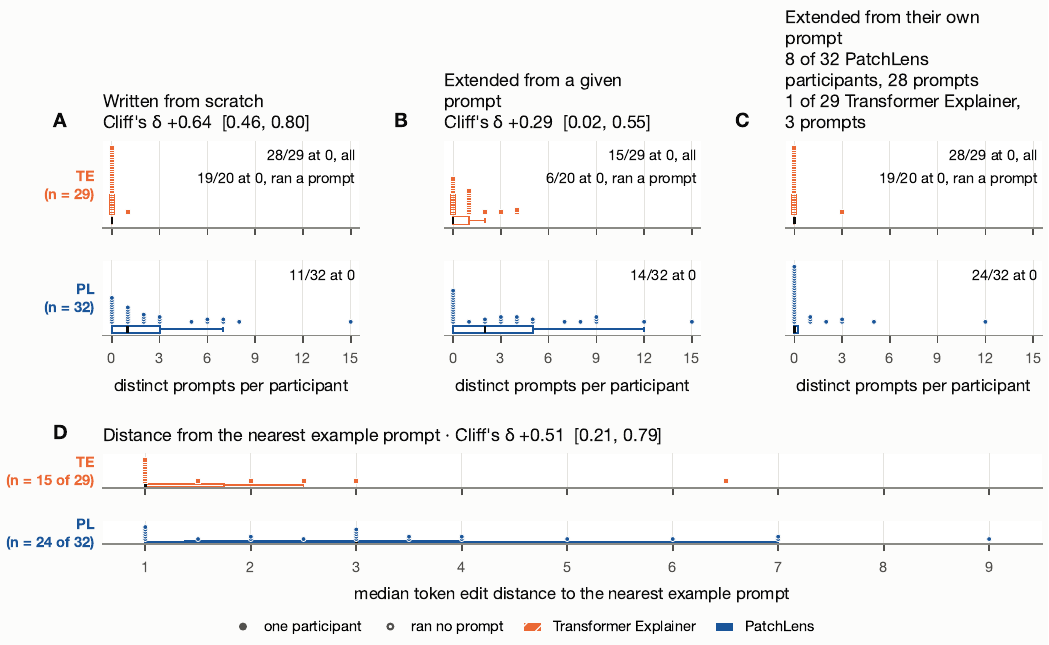}
    \caption{PatchLens participants supplied more of their own prompts, and prompts further from the supplied examples, than Transformer Explainer participants. Every analyzed participant is one marker, stacked at their count, over a box giving the median and IQR; participants who ran no prompt at all are drawn with an open marker, and Cliff's $\delta$ with its 95\% bootstrap interval is printed in each panel that carries one.
    (A) Prompts written from scratch.
    (B) Prompts extended from one of the tool's example prompts.
    (C) Prompts extended from a prompt the participant had written themselves.
    (D) Median token edit distance from each participant's own prompts to the nearest example prompt, over the 24 and 15 participants who supplied at least one.}
    \label{fig:prompts}
    \Description{Four sub-panels. Panels A, B, and C each show distinct prompts per participant on an axis from 0 to 15, with one row per arm: Transformer Explainer (n = 29) above and PatchLens (n = 32) below. Every analyzed participant is a marker stacked at their count, over a box giving the median and interquartile range, and participants who ran no prompt at all are drawn with an open marker. Panel A, prompts written from scratch, has 28 of 29 Transformer Explainer participants at zero against 11 of 32 under PatchLens, whose markers spread out to 15, with Cliff's delta of plus 0.64 and an interval from 0.46 to 0.80. Panel B, prompts extended from one of the tool's example prompts, has 15 of 29 and 14 of 32 at zero, so incidence is similar, but the PatchLens markers extend well past the Transformer Explainer maximum of 4, with Cliff's delta of plus 0.29 and an interval from 0.02 to 0.55. Panel C, prompts extended from a prompt the participant had written themselves, has 28 of 29 and 24 of 32 at zero. Panel D plots the median token edit distance from each participant's own prompts to the nearest example prompt, on an axis from 1 to 9, over the 15 Transformer Explainer and 24 PatchLens participants who supplied at least one. The Transformer Explainer markers concentrate at a distance of 1 while the PatchLens markers center on 3 and reach to 9, with Cliff's delta of plus 0.51 and an interval from 0.21 to 0.79.}
\end{figure}

PatchLens and Transformer Explainer both offer example prompts and accept the participant's own text. The PatchLens tutorial supplies eighteen prompts across its nine units, and Transformer Explainer opens on a default input a learner can replace or edit.
Participants ran 211 distinct prompts under PatchLens and 27 under Transformer Explainer, classified by provenance as in \S\ref{subsec:analysis} (Fig.~\ref{fig:prompts}).
Nine of the 29 Transformer Explainer participants ran no prompt, which the tool permits because it renders its default input without one. Every PatchLens participant ran at least one, because the lens requires a prompt to produce anything.

Twenty-one of 32 PatchLens participants wrote at least one prompt from scratch, against 1 of 29 using Transformer Explainer (Fig.~\ref{fig:prompts}.A), producing 78 distinct from-scratch prompts against one.
Comparable numbers in each arm extended one of the tool's examples (18/32 against 14/29), and the difference there is in volume rather than incidence. PatchLens has a median of 2 and a tail to 15 against a Transformer Explainer maximum of 4 (Fig.~\ref{fig:prompts}.B).
Eight PatchLens participants also extended a prompt they had written themselves, 28 prompts in total, against the same single Transformer Explainer participant with 3 (Fig.~\ref{fig:prompts}.C).

Transformer Explainer's own-prompt activity is almost entirely one behavior. Of its 27 prompts, 26 are appends to the tool's default sentence, and twelve of the fourteen participants who extended an example produced some version of that single stem.

PatchLens participants rebuilt the tutorial's probe forms with their own content.
One unit supplies ``Earlier I told you my favorite color. My favorite color is'' to show that the model has access only to the prompt.
Three participants rebuilt that probe: P08 ran \lstinline|Earlier I told you my favorite kind of pet. My favorite kind of pet is a| and \lstinline|What is my mother's name again? I told you yesterday,| and P16 ran \lstinline|My favorite color is Purple. Wait, what is my favorite color?|
Twelve participants ran factual completions modeled on the tutorial's landmark examples, such as \lstinline|the capital of australia is| (P14).
Fourteen built their own multi-shot arithmetic patterns after the unit demonstrating the form, and one carried its contradict-the-facts pattern from arithmetic into language, running \lstinline|Cats are made up of small mice.\n Dogs are made up of small| (P42).

The prompts participants supplied nonetheless departed from their closest examples. Among those who supplied at least one of their own, including appends (24/32 and 15/29), the median token edit distance to the nearest example is 3 under PatchLens against 1 under Transformer Explainer (Fig.~\ref{fig:prompts}.D).

\subsection{Experience (RQ3)}
\label{subsec:rq3-res}

Of the four experience measures, only mental demand separates the arms, and the open-ended responses locate the difficulty participants met in different places.

\subsubsection{Demand and Usability}
\label{subsubsec:res-demand}

\begin{figure}
    \centering
    \includegraphics[width=\linewidth]{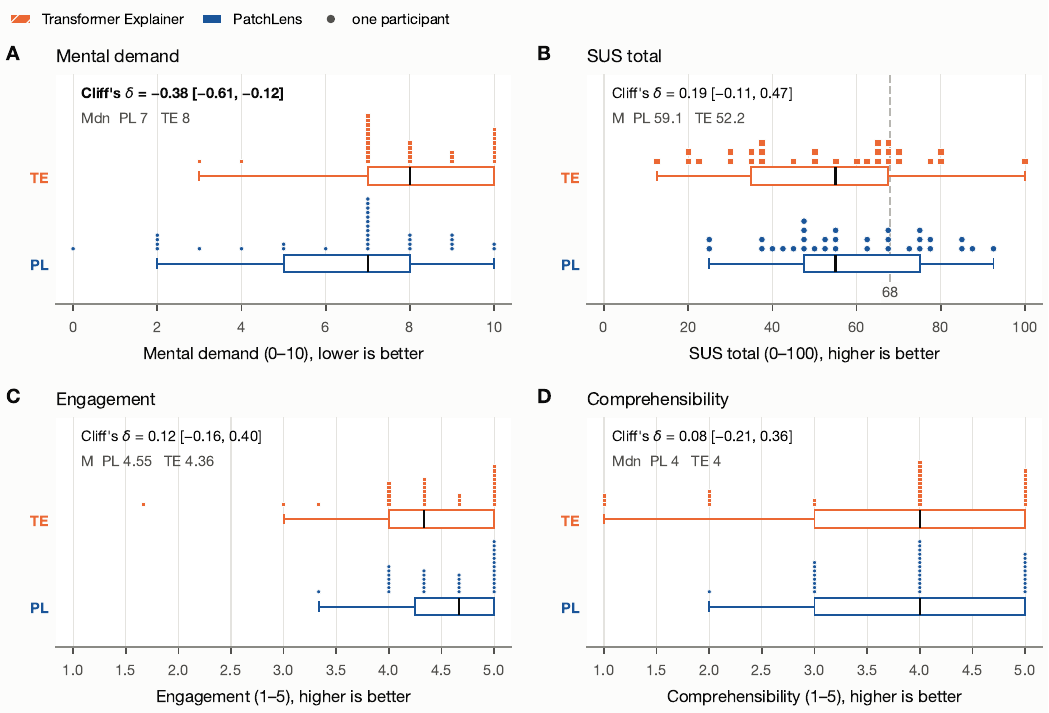}
    \caption{All four experience measures favor PatchLens, and only mental demand excludes zero. Each sub-panel shows a box (median, IQR, Tukey whiskers) with every participant overlaid as a dot; PatchLens is the bottom row throughout, and each sub-panel prints Cliff's $\delta$ with its 95\% bootstrap interval and the two arm summaries.
    (A) Self-rated mental demand, from 0 (no mental demand) to 10 (maximum mental demand).
    (B) SUS totals (0--100), with the 68-point benchmark as a dashed line.
    (C) Self-rated engagement, the mean of three 5-point agreement items: ``I enjoyed working through this material,'' ``I found this material interesting,'' and ``This material held my attention.''
    (D) Self-rated comprehensibility, 5-point agreement with ``I could follow the explanations without much confusion.''}
    \label{fig:usability}
    \Description{Four sub-panels, one per experience measure, each with two rows: Transformer Explainer above and PatchLens below. Each row is a box giving the median, interquartile range, and Tukey whiskers, with every participant overlaid as a dot, and each sub-panel carries Cliff's delta and its 95 percent interval as text. Panel A, self-rated mental demand from 0 to 10 with lower better, has medians of 8 and 7 and Cliff's delta of minus 0.38 with an interval from minus 0.61 to minus 0.12, the only interval in the figure that excludes zero; the PatchLens box extends further into the low end, with first quartiles of 7 and 5. Panel B, System Usability Scale totals from 0 to 100 with a dashed line at the 68-point benchmark, has means of 52.2 and 59.1 and Cliff's delta of 0.19 from minus 0.11 to 0.47; both medians fall left of the benchmark, and only the PatchLens upper quartile crosses it. Panel C, engagement from 1 to 5, has means of 4.36 and 4.55 and Cliff's delta of 0.12 from minus 0.16 to 0.40, with dots piled at the scale maximum in both arms. Panel D, comprehensibility from 1 to 5, has a median of 4 in both arms and Cliff's delta of 0.08 from minus 0.21 to 0.36. Across all four panels the PatchLens distribution is compressed toward the favorable end while the Transformer Explainer distribution carries a longer tail toward the unfavorable end, reaching the scale floor in panels C and D.}
\end{figure}

PatchLens imposed lower self-reported mental demand than Transformer Explainer
(Fig.~\ref{fig:usability}.A). The median is 7 against 8 (Cliff's $\delta = -0.38$, 95\% bootstrap
CI [$-0.61$, $-0.12$], Mann--Whitney $p = .009$).
It is the only experience measure whose interval excludes zero, and along with the prompt behavior of \S\ref{subsec:rq2-res} the only place in the study where an interval does.
The difference sits in the lower tail rather than in a whole-distribution shift. Ten of 32 PatchLens participants rated demand below 7, against 2 of 29 under Transformer Explainer.

On usability, engagement, and comprehensibility the intervals include zero (Fig.~\ref{fig:usability}).
Both arms fell below the SUS benchmark. PatchLens's 59.1 is 7.2 points above the 51.9 the pilot version scored on the same scale (\S\ref{sec:iteration}). The pilot ran with a different population in an instructor-led setting, so the change cannot be attributed to the revisions it motivated.

\subsubsection{What Participants Reported}
\label{subsubsec:res-openended}

\begin{figure}
    \centering
    \includegraphics[width=\linewidth]{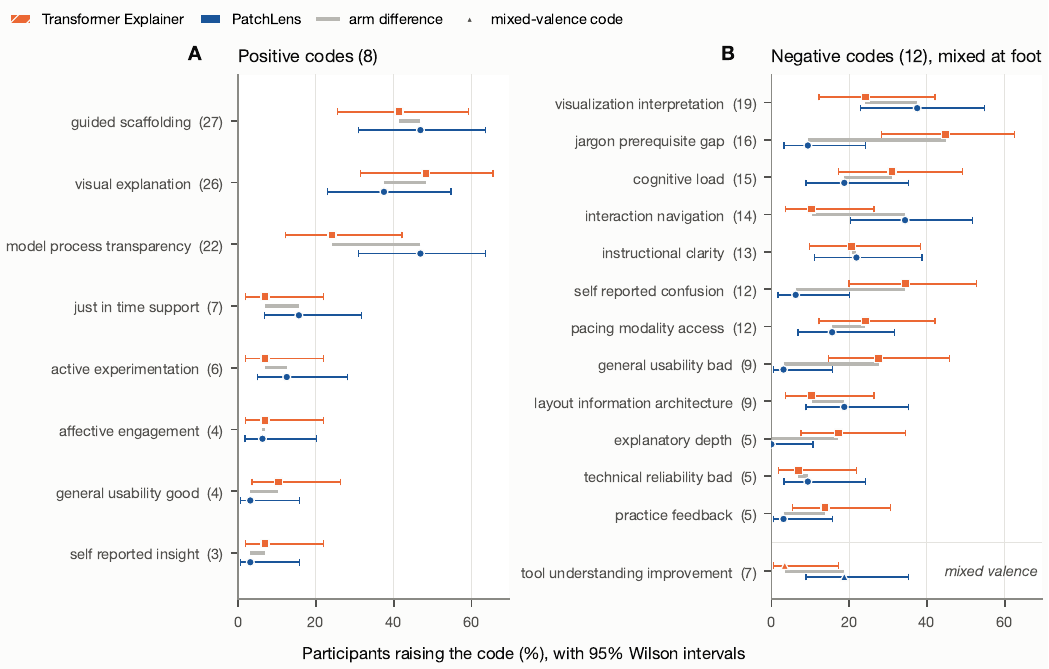}
    \caption{Prevalence of each qualitative code across the three open-ended prompts (most helpful aspects, most confusing aspects, and suggested improvements).
    (A) The eight positive codes and (B) the twelve negative codes, both sorted by overall prevalence, with the parenthesized number after each label giving the count of participants raising it. The one mixed-valence code, tool understanding improvement, sits at the foot of panel B with a triangular marker. Administrative codes are excluded: 14.6\% of PatchLens responses and 11.5\% of Transformer Explainer responses carried no substantive code.}
    \label{fig:qualcoderesults}
    \Description{Two panels of horizontal dot-and-interval plots, one row per qualitative code, sorted by overall prevalence, on a shared axis of the percentage of participants raising the code from 0 to 60 with 95 percent Wilson intervals. Each row carries one estimate per arm plus a grey bar spanning the difference between them, and the count of participants raising the code follows each label. Panel A holds the eight positive codes: guided scaffolding (27), visual explanation (26), model process transparency (22), just in time support (7), active experimentation (6), affective engagement (4), general usability good (4), and self reported insight (3). Guided scaffolding and visual explanation are the two most prevalent and the arms sit close together on both; the widest gap in the panel is model process transparency, where PatchLens is far to the right of Transformer Explainer. Panel B holds the twelve negative codes: visualization interpretation (19), jargon prerequisite gap (16), cognitive load (15), interaction navigation (14), instructional clarity (13), self reported confusion (12), pacing modality access (12), general usability bad (9), layout information architecture (9), practice feedback (5), explanatory depth (5), and technical reliability bad (5). The arms separate most on jargon prerequisite gap and self reported confusion, where Transformer Explainer is far to the right, and on interaction navigation, where PatchLens is. Below a rule at the foot of panel B sits the single mixed-valence code, tool understanding improvement (7), drawn with a triangular marker, on which PatchLens is to the right of Transformer Explainer. Intervals are wide throughout and overlap between arms on every code.}
\end{figure}

Participants from the two arms reported different kinds of difficulty (Fig.~\ref{fig:qualcoderesults}.B).
PatchLens participants mentioned difficulty using the interface more often (interaction navigation, 35\% against 10\%), while Transformer Explainer participants more often raised conceptual difficulty (jargon and prerequisite gaps, 10\% against 45\%) and explicit confusion (7\% against 34\%).
One PatchLens participant reported that ``the drag and drop source source [referring to activation patching], I spent a lot of time on that, and I don't think I got it to work'' (P08), while a Transformer Explainer participant pointed to ``the technical terms. Those were hard to grasp. and the graphic itself was too complex'' (P24).
Both arms reported being overloaded at comparable rates, with cognitive load raised by nine Transformer Explainer participants and six under PatchLens. Under PatchLens the obstacle was operating the tool and under Transformer Explainer it was the material.

Guided scaffolding and visual explanation were raised at comparable rates in both arms (47\% against 41\%, and 38\% against 48\%; Fig.~\ref{fig:qualcoderesults}.A).
The positive code that separated them was model process transparency, seeing what the model was doing with the input, raised by 47\% of PatchLens participants against 24\% under Transformer Explainer.
Participants described watching the prediction form rather than being told about it: ``I liked seeing all of the possibilities the AI ran through to create a response'' (P07), and ``the breakdown of how the AI thinks about each following word with a percentage for each was very useful and interesting'' (P42).

The interface difficulty was also reported as temporary more often under PatchLens.
Tool understanding improvement, the mixed-valence code recording an initial struggle that eased with use, was raised by 19\% of PatchLens participants against 3\% under Transformer Explainer. One described how ``interpreting the exact layer-by-layer transitions and tracking the shaded region took a little getting used to at first'' (P27).

\section{Discussion, Limitations, and Future Work}
\label{sec:discussion}
Participants with no formal background in AI ran logit lens and activation patching on a 70-billion-parameter model, and three findings follow.
Participants in both arms met difficulty but named different sources, interface navigation under PatchLens and textual content under Transformer Explainer.
That matters because the barrier to non-expert engagement with model internals has been attributed to the subject \cite{zhao_thinking_2025}, and our participants did not locate it there.
PatchLens participants also wrote their own prompts and Transformer Explainer participants almost never did, which suggests that what a tool has shown a learner how to ask is a design variable rather than a question of what the tool permits.
Self-reported mental demand was lower under PatchLens than under Transformer Explainer, which runs against the expectation that interactive tools raise cognitive load through the actions they require.

\subsection{Exposing Model Internals Shifts Difficulty from Material to Interface}
\label{subsec:difficulty}

Transformer Explainer participants wrote about the \textit{textual content}, naming unfamiliar terminology and background knowledge the material assumed, while PatchLens participants wrote about the \textit{interface controls}, naming grid navigation and patching.
\citet{cho_transformer_2026} reported Transformer Explainer less demanding than a blog post, which already ran against that expectation.
Our comparison presses further. PatchLens asks a learner to intervene on a model's internal state rather than edit an input or move a slider, and demand was lower again (\S\ref{subsubsec:res-demand}).

Where difficulty sat also bears on why interactive exploration has underperformed for language models in particular.
\citet{zhao_thinking_2025} attributed the decline they observed on the language-model topic to the models' probabilistic and less intuitive nature, and proposed that they may need scaffolding beyond surface-level interactive exploration (\S\ref{subsubsec:surfacelevel}).
Our formative workshop pointed to the representation rather than the technique. Participants struggled to map the conventional layer-by-position plot onto the residual stream, so the lens grid renders readouts as tokens in model coordinates (\ref{goal:logitlens}).
Some PatchLens participants still raised difficulty interpreting the grid, more often than Transformer Explainer participants did for their visualization (\S\ref{subsubsec:res-openended}). A representation can be legible in its parts and remain opaque as an object.
Participants nonetheless located difficulty in the interface rather than the material.
The decline \citet{zhao_thinking_2025} observed on that topic may therefore be a property of the representation a learner is given rather than of language models themselves. That makes internal state a specific candidate for the scaffolding they call for.

\subsection{A Revealed Mechanism Adds a Surface to Learn}
\label{subsec:surface}

Seamful design holds that revelation is always selective and that concealment is itself a design act (\S\ref{subsec:seamful}). Our results add a narrower point: the display doing the revealing is something a learner must learn before it gives access to the mechanism.
A tool built to show a mechanism therefore meets the question \citet{ehsan_seamful_2024} pose one level down, about the display.
The pilot's report of crowding recurred in the main study as layout complaints (\S\ref{subsubsec:res-openended}), so the density we chose not to reduce (\S\ref{sec:iteration}) is a cost this design carries rather than resolves.

Two of PatchLens's own bounds make the same point: patching is confined to a single layer, position, and prompt pair, and readouts come from raw logit lens. Neither bound announces itself in the output.
Under PatchLens, self-rating exceeded measured accuracy on six of nine objectives, and the three objectives scored lowest carry the largest gaps (\S\ref{subsubsec:res-selfassess}).
That descriptive pattern is a self-directed version of the over-confidence \citet{kaur_interpreting_2020} reported in practitioners reading interpretability output and \citet{schuff_human_2022} in end-users reading saliency maps.
Stating what a single intervention cannot establish where the learner reads the result, rather than in a caveat met earlier, is a requirement this version only partly meets.

\subsection{Learners Author Prompts by Reusing Question Forms}
\label{subsec:prompts}

The prompts PatchLens participants entered sat further from the supplied examples, where Transformer Explainer's own-prompt activity was almost entirely appending to its default sentence (\S\ref{subsec:rq2-res}).
Where a tool's central object is the prompt, as for a technique that reads one layer of one position, interaction takes the form of prompt authorship.
Both tools accepted arbitrary input and only one was used that way, so whether learners explore appears to depend less on permission than on what the tool has already shown them how to ask.

The prompts PatchLens participants wrote were not arbitrary: they rebuilt the question forms the tutorial had demonstrated, on content they chose.
The guided activity's units are the concrete-goal assignments \citet{de_jong_scientific_1998} recommend for hypothesis generation (\ref{goal:scaffold}). The support did more than scaffold the task it specified: it supplied a template for tasks it did not.
\citet{zamfirescu-pereira_why_2023} found that non-experts prompting an LLM tended to generalize from single observations rather than vary their inputs in a controlled way.
Our participants varied theirs, and what they varied was the content of a supplied form. The shortfall may therefore be in having no template to vary against rather than in the disposition to try.
\citet{renkl_structuring_2003} describe learning as a transition from studying worked examples to solving problems independently. The independence our sequence produced was independence over subject matter, with the form of the question still supplied.

\subsection{Teaching the Mechanism Without the Field's Vocabulary}
\label{subsec:vocabulary}

Within an inconclusive knowledge comparison (\S\ref{sec:results}), one asymmetry bears on tool design.
PatchLens teaches \textit{token} and next-token prediction and almost none of the rest of the field's vocabulary, demonstrating its objectives through what a learner can watch happen.
Its participants raised jargon far less often than participants in the arm whose tutorial names embeddings, attention heads, weights, and sampling parameters (\S\ref{subsubsec:res-openended}).
The one objective where that choice cost something is measured by an item written in terms of the model's weights, which a PatchLens participant could have understood without recognizing (\S\ref{subsec:studydesign}).
PatchLens participants were nonetheless above chance on every objective (\S\ref{subsubsec:res-knowledge}), though the instrument cannot show whether the account those objectives describe was acquired in other words.

\subsection{Considerations for Revealing Model Internals to Non-Experts}
\label{subsec:considerations}

Our design varied many properties at once and no condition isolated them, so what follows are considerations a tool designer, educator, or interpretability researcher can weigh rather than guidelines.

\textbf{Project internal state into the vocabulary the learner already reads.}
Showing activations as ranked tokens removes the step where a scalar must be decoded into a claim about the model.
The difficulty shifts from content to interface: a readable cell does not make a readable grid.

\textbf{Write guided activities whose question forms generalize.}
Because learners repurpose the question forms an activity supplies (\S\ref{subsec:prompts}), the choice of question form is a decision about what a learner explores unsupervised.
\citet{kawabe_technical_2024} reported both potential and limits in what non-experts took from freely exploring an interactive ML system.
A question form is a small enough unit to hand over. It gives a learner something to vary and a reason to expect one answer rather than another, which is agency over the hypothesis.

\textbf{Ask which of the field's terms a learner actually needs.}
PatchLens taught two of the field's terms and its participants were above chance on every objective (\S\ref{subsec:vocabulary}).
Which terms a non-expert needs to hold a working account of how a reply was produced is therefore open, and a tool designer choosing a vocabulary is answering it implicitly.
Deciding it explicitly, before the curriculum is written, also settles what an assessment of that curriculum may assume.

\textbf{Design the return trip to the deployed system, not only the departure from it.}
That the model under inspection is not the assistant a learner has used is a well understood requirement, and formative participants doubted the technique applied to their own questions when it was not met (\S\ref{sec:formative}).
The harder half is the return.
A learner leaves with an account of a model and returns to an application built around one, where a system prompt, retrieval, and an interface sit between the text they type and the weights they inspected.
Carrying a lesson across that boundary deserves as much design attention as setting the expectation did.

\subsection{Limitations and Future Work}
\label{subsec:limitations}

Knowledge was assessed only after tool use, so nothing here concerns learning within a participant, and the smallest difference the design could detect exceeded the difference observed.
Our instrument also inherits the vocabulary of the curriculum written alongside it, so an item phrased in one condition's terms cannot separate a learner who did not acquire a concept from one who acquired it in other words.
Where our estimate favors the comparison arm, the baseline imbalances and differential attrition (\S\ref{sec:results}) are candidate explanations alongside a difference in what the tools taught.
Post measures were applied immediately after use, so this study speaks neither to retention nor to transfer back to a deployed assistant.
The design evidence is narrower still. Both the formative workshop and the pilot drew on philosophy graduate students at a single summer school, so the challenges and goals were shaped by one population and tested on another.
Finally, the qualitative findings rest on brief answers to three open questions, so where participants located their difficulty is a difference in what they chose to write about rather than a measured effect.

Those bounds raise two questions.
The first is whether an architecture explanation and an intervention tool are the right pair.
We set them against each other because no tool combines them, and the difficulties participants reported were close to complementary.
A sequenced design, in which learners meet the architecture before intervening on it, would test whether the vocabulary one tool teaches makes the other's readouts interpretable. A within-subjects arrangement would let the same learner report which order helped.

The second is what a mechanism-focused tool changes about subsequent use, and the evidence is in conflict.
\citet{puppart_short-term_2025} found that a short intervention explaining a model's mechanism and limitations did not reduce adoption of incorrect recommendations, and increased the rate at which correct ones were ignored.
\citet{clerc_teaching_2026} found that a two-hour workshop improved query reformulation and accuracy in judging responses, and \citet{vu_student_2026} found that inoculation raised verification behavior specifically.
Our post-test could not detect which of those outcomes PatchLens moves.
The one self-report that moved was the item written at the grain of the intervention, while a general AI self-efficacy scale did not, so measures are worth matching in specificity to what a tool teaches.
Durability runs underneath both questions. We measured immediately after a single session, as \citet{cho_transformer_2026} also did. That matters more for mechanism knowledge than for procedural skill, because a learner returns not to the tool but to a chat interface that displays none of what they were shown.

\section{Conclusion}
\label{sec:conclusion}
Mechanistic interpretability can teach, not only answer research questions.
The shift is more than simplifying a research interface.
Both techniques report their output in the model's own vocabulary, so an intervention manipulates words a learner has already read rather than edits a hidden vector.
Each intervention also arrives with a question worth asking, which learners extended to their own experiments.
Our comparison was inconclusive on post-test knowledge, but an intervention a learner runs on a prompt they wrote is a different educational object from a result they are shown.
That is what we most encourage others to build on.

Interventions on LLMs are not free to build, but need not be built from scratch each time: guided material in PatchLens is authored as data rather than code. We expect the interpretability and AI education communities to find further ways to make model internals alterable by non-experts, not only visible to them.

\bibliographystyle{ACM-Reference-Format}
\bibliography{ref}

\appendix

\section{Knowledge Assessment}
\label{app:ka}
\begin{longtable}{@{}p{0.10\linewidth}p{0.86\linewidth}@{}}
  \caption{Full text of the knowledge-assessment items, administered once after
  tool use and identical across both conditions. LO tags refer to
  Table~\ref{tab:los}. Question order and answer order were both randomized in
  Qualtrics, so the letters below are presentational only and were not shown to
  participants; the keyed-correct response is marked with $^{\dagger}$ and was
  unmarked when fielded. All items were forced-response, back navigation was
  disabled within the block, and both question order and question answer order were randomized. Item text is participant-facing stimulus material
  and is reproduced verbatim, so it retains the double quotation marks and the
  abbreviation \emph{app} for \emph{application} that the paper's prose does
  not use.}
  \Description{A two-column table continuing across several pages. The left
  column gives the item label, K1 through K13, each tagged with the learning
  objective it measures. The right column gives the question stem followed by
  its four response options, labeled (a) through (d), with a dagger marking the
  keyed-correct option. The full text of every stem and option appears in the
  table body and is read in order by assistive technology. Items are ordered by
  item number rather than by objective, so the objectives appear in the order
  LO2, LO2, LO3, LO4, LO4, LO4, LO5, LO7, LO8, LO9, LO1, LO6, LO8. The
  keyed-correct option is (c) for K1, K3, K4, K7, K10, and K11; (a) for K2; and
  (b) for K5, K6, K8, K9, K12, and K13.} \\
  \label{tab:instrument} \\
  \toprule
  \textbf{Item} & \textbf{Question and response options} \\
  \midrule
  \endfirsthead
  \multicolumn{2}{@{}l}{\small\itshape Table~\ref{tab:instrument} continued from previous page} \\
  \toprule
  \textbf{Item} & \textbf{Question and response options} \\
  \midrule
  \endhead
  \midrule
  \multicolumn{2}{r@{}}{\small\itshape continued on next page} \\
  \endfoot
  \bottomrule
  \endlastfoot
  \textbf{K1} \;(LO2) & An AI chatbot (e.g., ChatGPT, Copilot, Gemini) uses a language model to write its replies. Which best describes how that model produces text? \\[2pt]
   & (a)~The model predicts which pre-written response best matches your message, then returns it. \\
   & (b)~The model works out all of the reply's tokens internally, then shows the finished text all at once. \\
   & (c)$^{\dagger}$~The model produces one token at a time, each predicted from the text so far. \\
   & (d)~The model produces one sentence at a time, each predicted from a list of candidates. \\
 
  \midrule
  \textbf{K2} \;(LO2) & When a chatbot (e.g., ChatGPT, Copilot, Gemini) writes a reply, how does the model it uses decide what text to produce? \\[2pt]
   & (a)$^{\dagger}$~The model generates responses based on patterns it learned from large amounts of text. \\
   & (b)~The model works like a search engine, generating each answer from information it pulls from the Internet. \\
   & (c)~The model stores its training data and returns exact answers from it. \\
   & (d)~The model follows hand-written rules that experts programmed for answering questions. \\
 
  \midrule
  \textbf{K3} \;(LO3) & A chatbot is producing a four-sentence answer to your question. Which best describes how the model it uses produces those four sentences? \\[2pt]
   & (a)~The model predicts all of the answer's tokens internally before any of them are shown, then shows the finished four sentences as one block. \\
   & (b)~The model predicts each of the four sentences as a whole, choosing one complete sentence at a time from a ranked list of candidates before showing it. \\
   & (c)$^{\dagger}$~The model produces the answer one token at a time. After each token, it runs again with everything so far as input to decide the next one. \\
   & (d)~The model produces the answer one token at a time. Each token is predicted from your question alone, not from the sentences it has already written. \\
 
  \midrule
  \textbf{K4} \;(LO4) & You have just sent another message in a long back-and-forth conversation with a chatbot. As the model predicts each next token of the reply, which text does it take as input? \\[2pt]
   & (a)~Only the most recent tokens. Earlier parts no longer affect the prediction. \\
   & (b)~Only your latest message. The model's earlier replies are not fed back in. \\
   & (c)$^{\dagger}$~All of the conversation so far, both your messages and the model's replies. \\
   & (d)~A summary the model keeps of the conversation so far, rather than the actual earlier text. \\
 
  \midrule
  \textbf{K5} \;(LO4) & Some chatbots have a ``memory'' feature that lets them use things from your past conversations. When the chatbot uses something it ``remembers,'' how does that past information actually reach the model? \\[2pt]
   & (a)~Your past conversations stay inside the model itself, so the information is there whenever you come back. \\
   & (b)$^{\dagger}$~The chatbot app adds the relevant past text into the input for this conversation, and the model itself is unchanged from one conversation to the next. \\
   & (c)~The model periodically retrains itself on your conversations, so that what you told it earlier becomes part of its knowledge and does not need repeating. \\
   & (d)~The model keeps a private log of past chats that it consults separately from the current input. \\
 
  \midrule
  \textbf{K6} \;(LO4) & You use a chatbot that has no memory features turned on. You start a brand-new conversation. As the model generates its first reply, what does it take as input? \\[2pt]
   & (a)~Everything you have said in all previous conversations. The model has also built a profile of your interests over time. \\
   & (b)$^{\dagger}$~Only the message you just sent, plus any system instructions the chatbot app adds. Earlier conversations are not included. \\
   & (c)~Your conversation history from the past 30 days, with anything older than that no longer being included. \\
   & (d)~Whatever the model judges most relevant from your past conversations, retrieved from the long-term store that it maintains for each user. \\
 
  \midrule
  \textbf{K7} \;(LO5) & A user asks a chatbot to classify movie reviews as positive or negative. Before asking about a new review, the user includes three reviews along with their correct labels in the same prompt. Compared to asking with no examples, what changes inside the model for this request? \\[2pt]
   & (a)~The model briefly retrains itself on the three labeled examples, adjusting its internal weights before it predicts the new label. \\
   & (b)~The model's weights are permanently updated to reflect the labels in the three examples, and stay updated for all future users' predictions. \\
   & (c)$^{\dagger}$~The model takes the three examples as part of the input it uses to predict the next token, and its weights are not changed. \\
   & (d)~The model searches its training data for the three reviews and copies whichever stored label is closest, so it does not predict a new one. \\
 
  \midrule
  \textbf{K8} \;(LO7) & A chatbot that cannot search the web or use other lookup tools confidently states a ``fact'' that turns out to be false, such as citing a court case that does not exist. Which best explains how the model it uses generated this false statement? \\[2pt]
   & (a)~The model retrieved the wrong entry from its stored database of facts. \\
   & (b)$^{\dagger}$~The model produced it by predicting likely text in the same way it produces true statements. \\
   & (c)~The model learned that false fact from its training data and predicted it back as it appeared there. \\
   & (d)~The model hit a temporary glitch and output text it normally would not. \\
 
  \midrule
  \textbf{K9} \;(LO8) & A language model writes: ``I am 99\% sure that the Eiffel Tower is in Berlin.'' What is the relationship between the expressed confidence (``99\% sure'') and how likely the statement is to be correct? \\[2pt]
   & (a)~The ``99\%'' is the model's internally calibrated probability that the statement is true, computed by a confidence-estimation step that runs before it predicts any output tokens. \\
   & (b)$^{\dagger}$~The ``99\%'' is generated text, predicted by the same process as the rest of the sentence, and it appears here next to a wrong fact. \\
   & (c)~The ``99\%'' reflects the model's measured average accuracy on geography questions of this kind, calculated over its training data before it generates the answer. \\
   & (d)~The ``99\%'' indicates that the model checked its training data before predicting the answer, and 99\% of the sources it found agreed with the statement. \\
 
  \midrule
  \textbf{K10} \;(LO9) & A chatbot that cannot search the web or use other lookup tools answers a question, and its answer turns out to be correct. Which best describes what the model it uses did? \\[2pt]
   & (a)~The model looked up the answer in an internal database of facts and returned the stored value, so it did not need to generate one. \\
   & (b)~The model predicted which stored question in its training data was closest, then returned the answer saved alongside it. \\
   & (c)$^{\dagger}$~The model generated the answer from patterns in its training, and the output matched a true fact but nothing was looked up. \\
   & (d)~The model predicted the answer by applying stored facts and rules it memorized during training to reach the conclusion. \\
 
  \midrule
  \textbf{K11} \;(LO1) & Which of the following is the clearest example of generative AI (as opposed to other kinds of AI)? \\[2pt]
   & (a)~A spam filter in your email that decides whether an incoming message belongs in your inbox or your junk folder. \\
   & (b)~A navigation app that finds the fastest driving route by checking current traffic against a map. \\
   & (c)$^{\dagger}$~A chatbot that writes a brand-new paragraph of text about whatever topic you describe to it. \\
   & (d)~A face-unlock feature on a phone that checks whether the person in front of the camera matches the owner's stored face. \\
 
  \midrule
  \textbf{K12} \;(LO6) & Why do the models that AI chatbots use have a maximum context length (a limit on how much text they can take in at once)? \\[2pt]
   & (a)~The companies that run chatbots cap how many tokens you can send to keep response times fast. The underlying models themselves have no maximum length. \\
   & (b)$^{\dagger}$~Each extra token increases the memory the model must hold and the calculation it must run, so the model itself has a limit. \\
   & (c)~The limit prevents users from sending more tokens than the total amount of text the model was trained on. \\
   & (d)~Longer inputs cause the model to predict off-topic and repetitive text, so a maximum length keeps responses focused. \\
 
  \midrule
  \textbf{K13} \;(LO8) & A language model gives an answer. You then ask, ``Why did you give that answer?'' and it produces a coherent explanation of the reasoning behind that answer. What is that explanation? \\[2pt]
   & (a)~A faithful trace of the steps the model took internally before it generated its previous answer, not newly written for your question. \\
   & (b)$^{\dagger}$~Generated text, predicted the same way as any other answer, so it sounds plausible but is not read off a record of the model's internal steps. \\
   & (c)~A simplified summary of the computation the model actually ran when it generated the previous answer, translated from its numerical representation into ordinary language. \\
   & (d)~A retrieval from an internal log the model keeps of the tokens it produced and why it chose them. \\
 
\end{longtable}

\section{Qualitative Codebook}
\label{app:codebook}

Table~\ref{tab:codes} gives the full code scheme applied to the 183 open-ended responses, with definitions, per-arm prevalence, and per-code agreement. Omnibus agreement figures are reported in \S\ref{subsec:analysis}.

\begin{table*}[t]
  \centering
  \caption{The 23-code scheme. Prevalence is the percentage of participants raising the code \emph{anywhere} across the three prompts (PL $n = 32$, TE $n = 29$); valence is a fixed property of the code, so pooling across prompts is the well-defined participant-level measure and no frequency is broken out by prompt. $\kappa$ is per-code Cohen's $\kappa$ on the two coders' \emph{independent} sets over 183 responses, and is unstable for codes with few instances; \texttt{Final} is consensus after discussion, which resolved 107 of 183 rows, and every prevalence uses the consensus codes. Omnibus agreement is reported in \S\ref{subsec:analysis}. Two documented recodes: \texttt{TECHNICAL\_RELIABILITY} was renamed \texttt{TECHNICAL\_RELIABILITY\_BAD} for symmetry with the usability pair, and \texttt{TECHNICAL\_RELIABILITY\_GOOD} --- one instance, no codebook definition, created during resolution --- was absorbed into \texttt{GENERAL\_USABILITY\_GOOD}. No significance test was run on any code frequency.}
  \Description{A table of the 23 qualitative codes grouped by polarity: positive, mixed, negative, then administrative. For each code the columns give its definition, its polarity, the percentage of participants in each arm who raised it, and per-code Cohen's kappa. Prevalence is pooled across the three open-response prompts and the denominators are 32 and 29 participants.}
  \label{tab:codes}
  \footnotesize
  \setlength{\tabcolsep}{4pt}
  \begin{tabular}{@{}l p{0.42\linewidth} l r r r@{}}
    \toprule
    \textbf{Code} & \textbf{Definition} & \textbf{Polarity} & \textbf{PL \%} & \textbf{TE \%} & \textbf{$\kappa$} \\
    \midrule
    \texttt{ACTIVE\_EXPERIMENTATION} & Open-ended manipulation or exploration to observe what changes. & positive & 12 & 7 & 1.00 \\
    \texttt{AFFECTIVE\_ENGAGEMENT} & Shares personal fascination or engagement. & positive & 6 & 7 & 0.89 \\
    \texttt{GENERAL\_USABILITY\_GOOD} & Positive comment on general usability without naming a specific interface element. & positive & 3 & 10 & 0.12 \\
    \texttt{GUIDED\_SCAFFOLDING} & Walkthroughs, sequencing, modeled examples, or step-by-step instruction. & positive & 47 & 41 & 0.62 \\
    \texttt{JUST\_IN\_TIME\_SUPPORT} & Hints, labels, definitions, pointers, or explanations at the point of need. & positive & 16 & 7 & 0.34 \\
    \texttt{MODEL\_PROCESS\_TRANSPARENCY} & Records exposure of model behavior, without necessarily converting the observation into a claim of understanding. & positive & 47 & 24 & 0.86 \\
    \texttt{SELF\_REPORTED\_INSIGHT} & An explicit claim of perceived understanding; not measured learning. & positive & 3 & 7 & 0.80 \\
    \texttt{VISUAL\_EXPLANATION} & Visual, interactive representations that make model behavior or data visible. & positive & 38 & 48 & 0.71 \\
    \addlinespace
    \texttt{TOOL\_UNDERSTANDING\_IMPROVEMENT} & Struggles to start, but reports that use became clearer or easier over time. & mixed & 19 & 3 & 0.00 \\
    \addlinespace
    \texttt{COGNITIVE\_LOAD} & Explicit overload from information volume, density, complexity, or simultaneous elements. & negative & 19 & 31 & 0.68 \\
    \texttt{EXPLANATORY\_DEPTH} & Explanations described as too shallow, or requests for more substantive detail. & negative & 0 & 17 & 0.66 \\
    \texttt{GENERAL\_USABILITY\_BAD} & Negative comment on general usability without naming a specific interface element. & negative & 3 & 28 & 0.44 \\
    \texttt{INSTRUCTIONAL\_CLARITY} & Requests improvements to the clarity of explanatory text. & negative & 22 & 21 & 0.21 \\
    \texttt{INTERACTION\_NAVIGATION} & Difficulty performing or locating an interface action. & negative & 34 & 10 & 0.57 \\
    \texttt{JARGON\_PREREQUISITE\_GAP} & Unfamiliar terminology, mathematics, formulas, or assumed background knowledge. & negative & 9 & 45 & 0.65 \\
    \texttt{LAYOUT\_INFORMATION\_ARCHITECTURE} & Clutter, competing interface elements, unclear grouping, or poor organization. & negative & 19 & 10 & 0.30 \\
    \texttt{PACING\_MODALITY\_ACCESS} & Time or pacing problems, requested learning formats, or an explicitly reported access need. & negative & 16 & 24 & 0.48 \\
    \texttt{PRACTICE\_FEEDBACK} & Structured practice, review questions, or correctness feedback. & negative & 3 & 14 & 0.33 \\
    \texttt{SELF\_REPORTED\_CONFUSION} & An explicit claim of confusion or misunderstanding. & negative & 6 & 34 & 0.49 \\
    \texttt{TECHNICAL\_RELIABILITY\_BAD} & Errors, loading failures, slowness, malfunction, or instructions perceived as incorrect. & negative & 9 & 7 & 0.83 \\
    \texttt{VISUALIZATION\_INTERPRETATION} & Difficulty understanding what displayed rows, layers, cells, tokens, or values mean. & negative & 38 & 24 & 0.52 \\
    \addlinespace
    \texttt{MISSING} & No response was supplied. & administrative & 9 & 3 & 1.00 \\
    \texttt{NO\_SUBSTANTIVE\_CODE} & An explicit absence of a suggestion or issue, rather than missing data. & administrative & 31 & 28 & 0.46 \\
    \bottomrule
  \end{tabular}
\end{table*}


\end{document}